\documentclass[12pt,journal,a4paper,draftcls,oneside,onecolumn]{IEEEtran}

\usepackage{times} 
\usepackage{graphicx}
\usepackage{cite}
\usepackage{citesort}
\usepackage{color}
\usepackage{psfrag}
\usepackage{subfigure}
\usepackage{amssymb}
\usepackage{epsfig}
\usepackage{pifont}
\usepackage{amsmath}
\usepackage{array}
\usepackage{multicol}
\usepackage[T1]{fontenc}
\usepackage[ruled, lined, linesnumbered, commentsnumbered, longend]{algorithm2e}
\usepackage{algorithmic}
\usepackage{comment}

\renewcommand{\baselinestretch}{1.26} 
\makeatletter
\def\ifundefined{\@ifundefined}
\makeatother 

\begin{document}


\title{{User-Centric Cell-Free (UCCF) Wireless Systems: Principles and Optimization}
\author{Lie-Liang Yang}
\thanks{L.-L. Yang is with the School of Electronics and Computer Science, University of Southampton, SO17 1BJ, UK. (E-mail: lly@ecs.soton.ac.uk). This document is a chapter in the book: L.-L. Yang, J. Shi, K.-T. Feng, L.-H. Shen, S.-H. Wu and T.-S. Lee, Resource Optimization in Wireless Communications: Fundamentals, Algorithms and Applications, Academic Press, USA (to be published in 2024).}}


\maketitle

\begin{abstract}
User-centric cell-free (UCCF) wireless networks have a range of
distinguished characteristics, which can be exploited for meeting some
challenges that the conventional cellular systems are hard to. This
chapter is devoted to delivering the fundamentals of wireless
communications in UCCF systems, including channel modeling and
estimation, uplink (UL) detection, downlink (DL) transmission, and
resource optimization. Specifically, the advantages of cell-free
networking are examined in contrast to the conventional celluar
systems.  The global and location-aware distributed UL detection are
explored in the principles of minimum mean-square error (MMSE) and
brief propagation. Correspondingly, the global and distributed DL
transmission schemes are designed based on the MMSE precoding. The
optimization of both UL and DL is analyzed with respect to system
design and resource-allocation. Furthermore, some challenges for the
implementation of UCCF systems in practice are identified and
analyzed.
\end{abstract}


\section{Introduction}\label{section-6G-4.1}
\begin{figure}[th]
  \begin{center} 
   \includegraphics[width=0.7\linewidth]{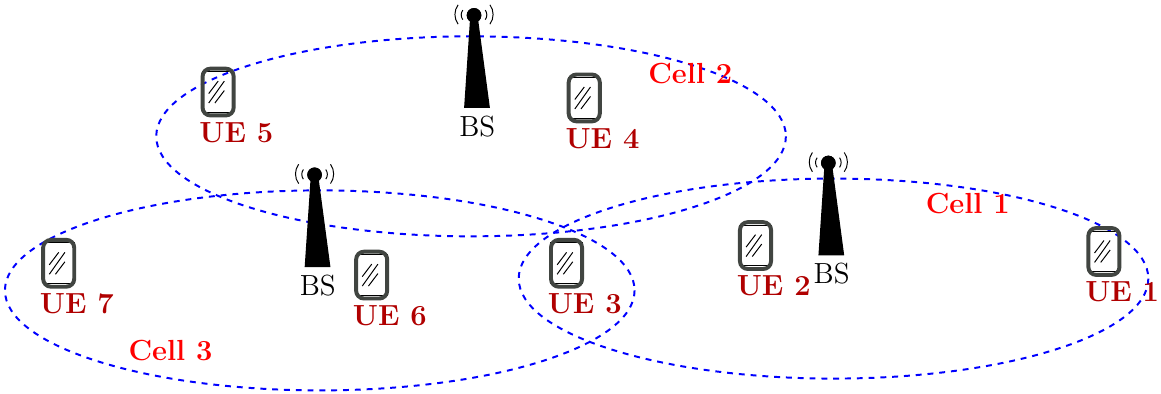}
   \end{center} 
  \caption{Illustration of cellular network structure.}
  \label{figure-Cellular-concept}
\end{figure}

In wireless communications, cellular networking\index{Cellular
  network} has been a major breakthrough in solving the problems of
spectral congestion and user capacity. Conceptually, a cellular
network has the structure as shown in
Fig.~\ref{figure-Cellular-concept}. A region covered by a cellular
system is divided into cells. Each cell is centred with a base-station
(BS) that is responsible for the control and signal transmissions of
the users, or user equipments (UEs), within the cell. In cellular
systems, each BS is allocated a fraction of the total number of
channels available to the entire system, and neighbouring BSs are
assigned different groups of channels to decrease the interference
between them. Nonetheless, the total available channels are
distributed throughout the geographic region and can be reused as many
times as necessary. Owing to this spectrum reuse, a cellular system is
capable of offering very high capacity by using a limited amount of
spectrum resource. In a given region, a higher capacity can be
obtained by dividing the region into more cells of smaller size.
\begin{figure}[th]
  \begin{center} 
   \includegraphics[width=0.75\linewidth]{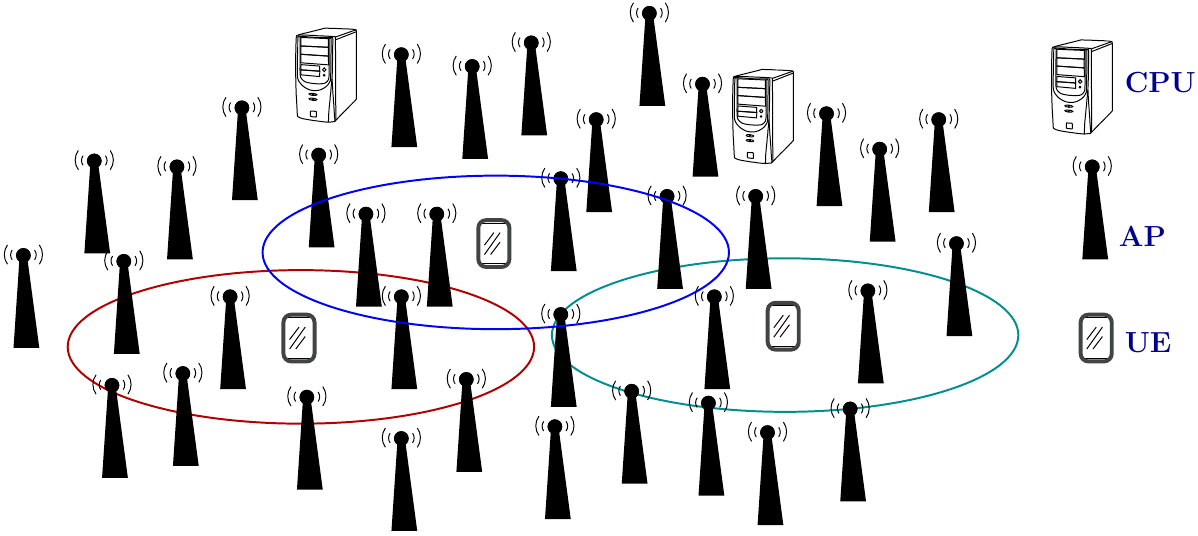}
   \end{center} 
  \caption{Illustration of UCCF network structure.}
  \label{figure-UCCF-concept}
\end{figure}

However, when wireless communications enters the generations that a
system's performance is not just measured by the performance, such as
capacity, reliability, etc., of itself, but mainly measured by the
qualities of services (QoS) demanded by users, the conventional
cellular networks encounter some deficiencies for providing services
in some applications. For example, in cellular systems, the cell-edge
users\index{cell-edge user} far away from BS, such as UEs 1, 3, 5, 7
in Fig.~\ref{figure-Cellular-concept}, may have to endure the poor
channels and reduce their requirements for QoS. The far away users may
also suffer strong interference from the users close to BS, such as
UEs 2, 4, 6 in Fig.~\ref{figure-Cellular-concept}, resulting in the
near-far problem\index{Near-far problem}. Accordingly, BS may have to
limit the desire of the users close to BS for the best possible QoS,
so as to protect the cell-edge users.  Furthermore, the spectrum
reuse\index{Spectrum reuse} with a reuse factor\footnote{Spectrum
  reuse factor is defined as the rate at which the same spectrum can
  be used in a network. A reuse factor $1/N$ means that a cluster of
  $N$ cells cannot use the same spectrum to transmit
  signals.}\index{Spectrum reuse!factor} less than one limits the
exploitation of the precious frequency resource, reducing the
spectral-efficiency. Moreover, spectrum reuse generates inter-cell
interference\index{Inter-cell interference} or so-called co-channel
interference\index{Co-channel interference}, making the cell-edge
users, such as UE 3 in Fig.~\ref{figure-Cellular-concept}, vulnerable.

To circumvent the shortcomings of cellular systems presented in some
applications, the concept of user-centric cell-free
(UCCF)\index{User-centric cell-free (UCCF)} wireless communications
came in sight~\cite{1682819,4696029,9650567}. UCCF networks have the
structure as shown in Fig.~\ref{figure-UCCF-concept}. A UCCF network
may include one to several central processing units
(CPUs)\index{Central processing unit (CPU)}, and many distributed
access points (APs)\index{Access point (AP)} that are connected with
CPU(s) via fiber optics or dedicated radio resources. APs and CPU(s)
work together to serve the UEs in the network. A UCCF network may be a
heterogeneous network with the dense deployment of APs.  In contrast
to the cellular systems where BS is the center of a cell, in UCCF
networks, every user is the center of its virtual cell\footnote{The
  virtual cell of a user is defined as the certain area around the
  user.}\index{Virtual cell}, which sends signals to and receives
signals from the APs located in its virtual cell.

In comparison with cellular networks, UCCF networks have their unique
characteristics for meeting some challenges that cellular networks are
hard to. First, by deploying many distributed APs, instead of the
co-located antennas at BS, user (or system) capacity can be
significantly increased for given spectrum and antenna (space)
resources.  Specifically, as shown in \cite{1682819,4696029}, a
distributed antenna system with user-centric detection is capable of
supporting many more users than a corresponding cellular system with
co-located antennas at BS. In addition to the capacity benefit brought
by distributed APs, in UCCF systems, a spectrum reuse factor of one
can be implemented, which also engenders the capacity advantage of
UCCF systems over the conventional cellular systems.

In UCCF networks, APs are usually close to UEs. Hence, signals sent by
APs to UEs or by UEs to APs do not experience severe propagation
pathloss and shadowing\index{Large-scale
  fading!pathloss}\index{Large-scale fading!shadowing}. The chance of
having line-of-sight (LoS) transmission paths between APs and UEs is
high. Therefore, wireless communications in UCCF networks is high
energy-efficiency. In UCCF networks, any UE is a strong user in its
virtual cell, hence, there is no near-far problem\index{Near-far
  problem}, rendering that low-complexity detection is
near-optimum. consequently, in UCCF networks,
power-control\index{Power-control} is mainly QoS dependent, but not
for the near-far problem, as done in cellular
systems~\cite{Viterbi_book,book:lie-liang}.

There is no handover\index{Handover} as that in cellular systems
needed in UCCF networks. When a UE moves, its virtual cell also moves,
some current APs may move out of its virtual cell, while some new APs
are added in it. During the process, system can dynamically update the
APs serving the UE, according to the real-time signal measurement, or
simply, the prediction of the UE's movement. In UCCF networks, the
sizes of UEs' virtual cells can be dynamic and variant, being set
according to UEs' specific communications environments or/and their
QoS requirements. For example, a UE having a higher reliability or/and
rate requirement can have a virtual cell of bigger size than a UE
having a lower requirement for reliability or/and rate.

UCCF networks are high-robustness. APs can be arbitrary added to a
UCCF network to enhance its performance. Some random failures of APs
would not result in the catastrophic effect on communications. UCCF
networks are feasible for meeting the demands of various types of
services, e.g., high-rate, high-reliability, low-latency. The required
QoS can be met via setting an appropriate size for the virtual cell,
allocating the relevant spectrum and time resources, executing the
tailored signal processing, etc.

In UCCF networks, the signal processing for transmission and detection
is user location aware.  On uplink (UL) transmission, one UE's
transmitted signal is only received by a small subset of APs that are
around the UE. On downlink (DL) transmission, usually, only a few of
APs surrounding a UE are dedicated to transmitting signals to the
UE. As above-mentioned, this networking feature brings the advantages
of, such as, energy-efficiency, efficiency of resource usage,
etc. Moreover, it can be shown that the inputs and outputs of such a
UCCF network are related by a sparse graph\index{Sparse graph}, which
is beneficial to engaging the high-efficiency brief-propagation
algorithms\index{Brief-propagation} for attaining near-optimum
performance in signal detection~\cite{9220147}.

Additionally, UCCF networking may provide a promising method for the
implementation of secrecy communications at physical layer, which will
be explained in Section~\ref{section-6G-4.7}.

Naturally, there are challenges to meet in the design, optimization
and implementation of UCCF networks. Therefore, this chapter motivates
to address the fundamentals of UCCF systems and explain some possible
challenges. Specifically, in Section~\ref{section-6G-4.2}, the system
models for UCCF networks are discussed. Section~\ref{section-6G-4.3}
focuses on the channel modeling and channel estimation, while
Section~\ref{section-6G-4.4} provides some UE association methods. In
Section~\ref{section-6G-4.5}, the UL detection and optimization are
analyzed, while Section~\ref{section-6G-4.6} considers the DL
transmission and optimization. Finally, Section~\ref{section-6G-4.7}
summarizes the chapter and provides some concluding remarks.

\section{System Models for UCCF Wireless Networks}\label{section-6G-4.2}

A UCCF network has the structure as shown in
Fig.~\ref{figure-UCCF-concept}. As above-mentioned, in a UCCF network,
APs are geographically distributed in an area, which are connected via
fiber optics or dedicated radio resources to one or to multiple CPUs,
where network control functions and/or main signal processings are
implemented. In a UCCF network, each AP usually only serves a small
number of UEs, responsible for their signal receiving and
transmission. Each UE is usually associated with one to a few of
APs. The association\index{Association!user equipment} of UEs to APs
may be simply based on the physical distances between UEs and APs or
other measurements, as that to be discussed in
Section~\ref{section-6G-4.4}. Depended on the practical application
scenarios, a UE may employ one or several antennas or an antenna
array. Similarly, an AP may be equipped with one antenna, multiple
antennas or even several antenna arrays. Furthermore, in UCCF
networks, various duplex techniques between UL and DL, multiuser
multiplexing schemes and signaling methods may be
implemented~\cite{4696029,9582783,9931474,10153484,10310239,9650567}.

In this chapter, a UCCF network with baseband OFDM
signaling\index{Signaling!OFDM} is considered to analyze the UL
detection and DL preprocessing, as well as their related
optimization. Specifically, a UCCF network employing one CPU connected
with $M$ APs, which may be randomly or regularly distributed, is
assumed to support $K$ UEs. Each of APs and UEs is equipped with one
antenna for receiving or transmission. The number of subcarriers of
OFDM is denoted by $N$, whose indices form a set
$\mathcal{N}=\{1,2,\ldots,N\}$. After UE association, the set of UEs
associated with AP $m$ is expressed as $\mathcal{K}_m$ for
$m=1,2,\ldots,M$, and the set of APs monitoring the $k$th UE is
expressed as $\mathcal{M}_k$ for $k=1,2, \ldots,K$. Hence,
$\mathcal{M}=\mathcal{M}_1\cup \mathcal{M}_2\cdots\mathcal{M}_K$, and
$\mathcal{K}=\mathcal{K}_1\cup \mathcal{K}_2\cdots\mathcal{K}_M$,
where $\mathcal{M}=\{1,2,\ldots,M\}$ and
$\mathcal{K}=\{1,2,\ldots,K\}$. Both UL transmissions from UEs and DL
transmissions to UEs are assumed to be synchronous. Below let us first
consider the channel modeling and estimation in UCCF networks.

\section{Channel Modeling and Estimation}

In this section, the channel modeling for UCCF networks is first
provided. Then, the principles of channel estimation is explained.

\subsection{Channel Modeling}\label{subsection-6G-4.3.1}\label{section-6G-4.3}\index{Channel!modeling|(}

Consider a UCCF system where channels experience both the large-scale
propagation pathloss and shadowing\index{Large-scale
  fading!pathloss}\index{Large-scale fading!shadowing}, and the
small-scale fading\index{Small-scale fading}, the channel impulse
response (CIR)\index{Channel impulse response (CSI)} between UE $k$
and AP $m$ can be denoted as
\begin{align}\label{eq:UCCF-1}
\pmb{h}_{mk}=\sqrt{g_{mk}}\tilde{\pmb{h}}_{mk},~k\in\mathcal{K},m\in\mathcal{M}
\end{align}
where $\tilde{\pmb{h}}_{mk}\in\mathcal{C}^{L_{mk}\times 1}$ accounts
for the small-scale fading, $L_{mk}$ is the number of taps of CIR and
$E\left[\|\tilde{\pmb{h}}_{mk}\|^2\right]=1$. $\tilde{\pmb{h}}_{mk}$
varies relatively fast, with the coherence
period\index{Channel!coherence period} expressed as $\tau_c$. In
\eqref{eq:UCCF-1}, $g_{mk}$ models the large-scale propagation
pathloss and shadowing effect. In practice, $g_{mk}$ varies much
slower than $\tilde{\pmb{h}}_{mk}$. Hence, when averaging out the
effect of $\tilde{\pmb{h}}_{mk}$, the transmit power $P_{t}$ of UE $k$
and the receive power $P_r$ of AP $m$ from UE $k$ have the
relationship of $P_r=g_{mk}P_t$.

Typically, $g_{mk}$ can be modelled to obey the lognormal
distribution\index{Lognormal distribution}, with a PDF expressed
as~\cite{4696029,book-Simon-Alouini-2nd-Ed}
\begin{align}\label{eq:UCCF-2}
f_{g_{mk}}(x)=\frac{\xi}{\sqrt{2\pi}\sigma_{g}x}\exp\left[\frac{(10\log_{10}x-\mu(d_{mk}))^2}{2\sigma_g^2}\right],~x>0
\end{align}
where $\xi=10/\ln 10=4.3429$, $\mu(d_{mk})$ (dB) and $\sigma_g$ (dB)
are the mean and standard deviation of $10\log_{10}g_{mk}$,
respectively, $d_{mk}$ is the distance between UE $k$ and AP $m$. The
mean $\mu(d_{mk})$ (dB) accounts for the propagation pathloss. In
\cite{4696029}, the {\em double-slope} propagation pathloss model was
employed for performance evaluation, which is represented as\index{Large-scale
  fading!pathloss!double-slope}
\begin{align}\label{eq:UCCF-3}
\mu(d_{mk})=-10\log_{10}\left[d_{mk}^a\left(1+\frac{d_{mk}}{d_{Break}}\right)^b\right]
\end{align}
where $a$ is referred to as the basic pathloss exponent, which takes a
value of approximately 2, $b$ is the additional pathloss exponent,
which has a value ranging from 2 to 6, and $d_{Break}$ is referred
to as the break point of the propagation pathloss curve. 

Another propagation pathloss model applied in the research on
cell-free systems~\cite{8000355,9586055} is the {\em triple-slope}
model, which can be represented as~\cite{944859}\index{Large-scale
  fading!pathloss!triple-slope}
\begin{align}\label{eq:UCCF-4}
\mu(d_{mk})=\left\{\begin{array}{ll}
-L_P-35\log_{10}(d_{mk}), &\textrm{if $d_{mk}>d_1$}\\
-L_P-10\log_{10}\left(d_1^{1.5}d_{mk}^{2}\right), &\textrm{if $d_0<d_{mk}\leq d_1$}\\
-L_P-10\log_{10}\left(d_1^{1.5}d_{0}^{2}\right), &\textrm{if $d_{mk}\leq d_0$}\\
\end{array}\right.
\end{align}
for certain $d_0$ and $d_1$ values, where $L_P$ is given as
\begin{align}\label{eq:UCCF-5}
L_P=&46.3+33.9\log_{10}f-13.82\log_{10}h_{AP}\nonumber\\
&-[1.11\log_{10}f-0.7]h_{UE}+1.56\log_{10}f-0.8
\end{align}
with $f$ defined as the carrier frequency in MHz, $h_{AP}$ and
$h_{UE}$ the antenna heights of AP and UE.

\index{Channel!modeling|)}
\subsection{Channel Estimation}\label{subsection-6G-4.3.2}\index{Channel!estimation|(}

%
\begin{figure}[th]
  \begin{center} 
   \includegraphics[width=0.65\linewidth]{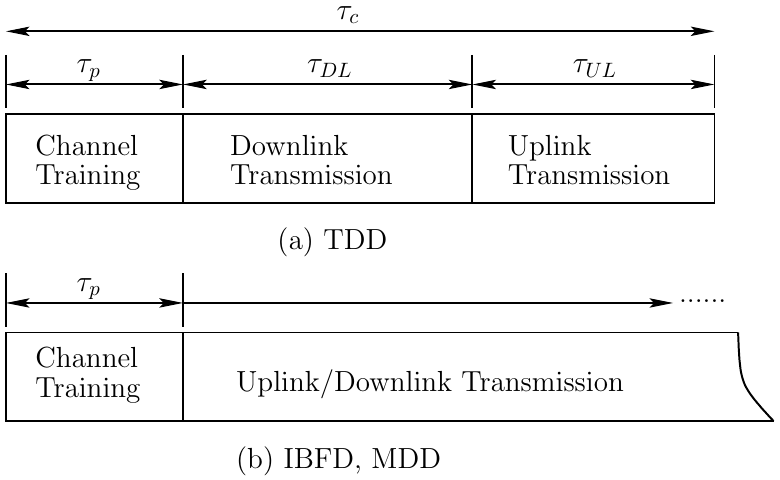}
   \end{center} 
  \caption{Frame structures for the UCCF systems with TDD, MDD, and
    IBFD, respectively.}
  \label{figure-UCCF-frame-structure}
\end{figure}

Assume a UCCF operational scenario, where channel's coherence time (or
period, duration)\index{Channel!coherence
  time}\index{Channel!coherence duration} is denoted as $\tau_c$, as
shown in Fig.~\ref{figure-UCCF-frame-structure}. When time-division
duplex (TDD)\index{Duplex!time-division (TDD)} is employed, signal
transmission over one coherence duration can be arranged as shown in
Fig.~\ref{figure-UCCF-frame-structure}(a). First, $\tau_p<\tau_c$ is
used for UL channel training to estimate channels. Then, relying on
the reciprocity between DL and UL channels, the estimated channel
state information (CSI) is first applied for DL signal transmission to
UEs, followed by using it for UL signal detection at APs or at CPU.

TDD is beneficial to UCCF systems for saving the overhead of CSI
acquisition, in comparison with frequency-division duplex
(FDD).\index{Duplex!frequency-division (FDD)} However, in some
applications, e.g., in fast time-varying communication environments,
TDD may be deficient. First, when a wireless channel becomes more time
variant, the channel's coherence duration $\tau_c$ becomes
shorter. Hence, when given the time duration $\tau_p$ required for
channel training, the duration left for DL/UL data transmission
becomes shorter, overall, resulting in a significant overhead increase
of channel acquisition. Second, the CSI estimated at the beginning of
a coherence duration becomes less accurate with time, yielding the
channel ageing problem~\cite{9582783,9903599}.\index{Channel ageing}
If channel varies fast, the CSI applied for DL transmission and,
especially, for UL detection may become outdated, resulting in
significant performance degradation.

To mitigate the channel aging problem and save the overhead consumed
for CSI acquisition under TDD, in-band full-duplex
(IBFD)~\cite{7024120}\index{Duplex!in-band full-duplex (IBFD)} may be
employed to enable the simultaneous DL and UL transmissions on the
same spectrum. Accordingly, the CSI for DL transmission and UL
detection can be obtained whenever needed, by appropriately inserting
UL pilots according to requirement or, simply, by employing the UL
decision-directed channel estimation,\index{Channel
  estimation!decision-directed} which estimates channels with the aid
of the reliably detected UL data. However, IBFD conflicts severe
self-interference (SI),\index{Self-interference (SI)} making channel
estimation hard or even impossible, if SI is not sufficiently
suppressed. Yet, the dilemma is that most efficient techniques for the
SI suppression in IBFD are relied on CSI.

To circumvent the dilemma of IBFD but avoiding the shortcomings of
TDD, multicarrier-division duplex
(MDD)~\cite{Lie-Liang-MC-CDMA-book,9582783,9184063,9477180,9903599}\index{Duplex!multicarrier-division
  (MDD)} may provide the solution. MDD belongs to an out-band
full-duplex scheme, where a portion of subcarriers in a spectrum are
assigned to support UL transmission, while the rest subcarriers are
used to support DL transmission. Hence, as TDD, MDD enjoys the DL/UL
reciprocity for CSI estimation, owing to the frequency-domain channel
correlation.  As IBFD, MDD possesses the properties of full-duplex,
allowing simultaneous DL and UL transmissions. Hence, as IBFD,
instantaneous CSI is available for DL transmission and UL detection in
the MDD-supported systems, and hence, there is no channel aging
problem. With regard to SI, in contrast to IBFD, which demands SI
suppression in propagation, analog and digital domains, MDD only
requires SI suppression in propagation and analog domains to protect
the received signals to linearly pass analog-to-digital converters
(ADCs). In this case, MDD enables the received signals to be free from
digital-domain SI~\cite{9184063}.

Hence, both IBFD- and MDD-aided systems may use the frame structure as
shown in Fig.~\ref{figure-UCCF-frame-structure}(b). During one session
of communication, which may be much longer than the channel's
coherence duration, channel training is only required at the beginning
of the transmission session. Then, as above-mentioned, CSI can be
updated whenever needed, with the aid of the appropriately inserted UL
pilots or the decision-directed channel estimation. Hence, a
significant overhead required for the channel acquisition by TDD can
be saved, while without experiencing the channel aging problem.

Below the channel estimation during the initial channel training
stage, as shown in Fig.~\ref{figure-UCCF-frame-structure}(a) and (b),
is analyzed. Assume that a pilot symbol block used by UE $k$ for
channel training is expressed as
$\pmb{S}_k\in\mathcal{C}^{N_k\times\tau_p}$, where $N_k$, $L_k\leq N_k\leq
N$, is the number of pilot symbols sent per OFDM symbol duration, and
$\tau_p$ is the number of OFDM symbol periods used for training. It is
assumed that $E[|\pmb{S}_k(i,j)|^2]=1$, and $\pmb{S}_k$ is known
to the APs that UE $k$ is associated with or/and to the CPU.

When pilot symbols are sent over the channels with the CIR expressed
as \eqref{eq:UCCF-1}, by following the principles of OFDM\index{OFDM}
in Chapter~{\bf (OFDM)}, the frequency-domain observations obtained by
AP $m$ (assume that UE $k$ is associated with AP $m$) can be expressed
as
\begin{align}\label{eq:UCCF-6}
\pmb{Y}_{m}=\sum_{k\in\mathcal{K}_m}\sqrt{P_k}\textrm{diag}(\pmb{F}_N\pmb{\Psi}_{mk}\pmb{h}_{mk})\pmb{\Phi}_{k}\pmb{S}_k+\pmb{J}_m+\pmb{N}_m,~m\in\mathcal{M}
\end{align}
where $\pmb{Y}_{m}\in\mathcal{C}^{N\times\tau_p}$,
$\pmb{N}_m\in\mathcal{C}^{N\times\tau_p}$ is the Gaussian noise
distributed with mean zero and a variance $\sigma^2/2$ per dimension,
and $\pmb{J}_m\in\mathcal{C}^{N\times\tau_p}$ is the possible
interference from the other UEs not associated with AP $m$, which can
be modelled to obey a Gaussian distribution with zero mean and a
variance $\sigma_J^2/2$ per dimension.  $P_k$ is the transmit power
per active subcarrier of UE $k$. $\pmb{F}_N$ is the non-normalized FFT
matrix satisfying
$\pmb{F}_N\pmb{F}_N^H=\pmb{F}_N^H\pmb{F}_N=N\pmb{I}_N$.
$\pmb{\Psi}_{mk}$ is a mapping matrix consisting of the first $L_k$
columns of identity matrix $\pmb{I}_N$, and $\pmb{\Phi}_{k}$ is a
mapping matrix, constructed by the $N_k$ columns of $\pmb{I}_N$ that
correspond to the $N_k$ subcarriers activated by UE $k$ to send pilot
symbols. $\pmb{\Psi}_{mk}$ and $\pmb{\Phi}_{k}$ are also known to AP
$m$ or/and CPU.  Finally, $\textrm{diag}(\pmb{a})$ yields a diagonal
matrix using vector $\pmb{a}$.

Upon vectorizing \eqref{eq:UCCF-6} and expressing
$\pmb{y}_m=\textrm{vec}(\pmb{Y}_{m})$,
$\pmb{n}_m=\textrm{vec}(\pmb{N}_m)$ and
$\pmb{j}_m=\textrm{vec}(\pmb{J}_m)$, which are $N\tau_p$-length
vectors, we have
\begin{align}\label{eq:UCCF-7}
\pmb{y}_m=&\sum_{k\in\mathcal{K}_m}\sqrt{P_k}\tilde{\pmb{S}}_k\pmb{F}_N\pmb{\Psi}_{mk}\pmb{h}_{mk}+\pmb{j}_m+\pmb{n}_m,\nonumber\\
=&\sum_{k\in\mathcal{K}_m}\pmb{A}_{mk}\pmb{h}_{mk}+\pmb{j}_m+\pmb{n}_m,~m=1,2,\ldots,M
\end{align}
where
$\tilde{\pmb{S}}_k=\left[\textrm{diag}(\pmb{\Phi}_{k}\pmb{s}_{k1}),\ldots,\textrm{diag}(\pmb{\Phi}_{k}\pmb{s}_{k\tau_p})\right]^T$
with $\pmb{s}_{ki}$ being the $i$th column of $\pmb{S}_k$, and for
simplicity,
$\pmb{A}_{mk}=\sqrt{P_k}\tilde{\pmb{S}}_k\pmb{F}_N\pmb{\Psi}_{mk}$ is
defined. Note that, $\tilde{\pmb{S}}_k$ is a $(N\tau_p\times N)$
matrix and $\pmb{A}_{mk}$ is $(N\tau_p\times L_k)$-dimensional.

The objective of channel training is to estimate $\pmb{h}_{mk}$ for
all $k\in\mathcal{K}$ and their associated APs, which may be carried
out locally at APs or at CPU. Assume that channel estimation is
executed locally at APs.  If pilot sequences are designed and assigned
to UEs to make the terms in $\{\pmb{A}_{mk}\}$ nearly orthogonal,
i.e., $\pmb{A}_{mk}^H\pmb{A}_{nl}$ gives an almost orthogonal matrix
for $k\neq l$, individual UE's channel can be estimated without
considering multiuser interference (MUI) mitigation. Specifically,
when the minimum mean-square-error (MMSE)\index{Minimum
  mean-square-error (MMSE)} relied channel
estimation~\cite{book-Steven-Kay-I} is employed, $\pmb{h}_{mk}$ can
be estimated as
\begin{align}\label{eq:UCCF-8}
\hat{\pmb{h}}_{mk}=\pmb{C}_{mk}^{-1}\pmb{Q}_{mk}^H\pmb{A}_{mk}^H\left[\pmb{A}_{mk}\pmb{Q}_{mk}\pmb{A}_{mk}^H+(\sigma_J^2+\sigma^2)\pmb{I}\right]^{-1}\pmb{y}_m
\end{align}
where $\pmb{Q}_{mk}=E\left[\pmb{h}_{mk}\pmb{h}_{mk}^H\right]$ is the
covariance matrix of $\pmb{h}_{mk}$, and $\pmb{C}_{mk}$ is introduced
to obtain an unbiased estimator, which is a diagonal matrix formed by
the diagonal elements of
$\pmb{Q}_{mk}^H\pmb{A}_{mk}^H\left[\pmb{A}_{mk}\pmb{Q}_h\pmb{A}_{mk}^H+(\sigma_J^2+\sigma^2)\pmb{I}\right]^{-1}\pmb{A}_{mk}$.

However, if MUI suppression is required during channel estimation, $\pmb{h}_{mk}$ can be estimated as
\begin{align}\label{eq:UCCF-9}
\hat{\pmb{h}}_{mk}=\pmb{C}_{mk}^{-1}\pmb{Q}_{mk}^H\pmb{A}_{mk}^H\left[\sum_{l\in\mathcal{K}_m}\pmb{A}_{ml}\pmb{Q}_{ml}\pmb{A}^H_{ml}+(\sigma_J^2+\sigma^2)\pmb{I}\right]^{-1}\pmb{y}_m
\end{align}

In \eqref{eq:UCCF-9}, the matrix in the bracket is constructed using
the pilots assigned to UEs, the mapping matrices used by UEs, and the
statistics about interference and noise. In the case that these
requirements cannot be satisfied, a matrix can be directly estimated
from \eqref{eq:UCCF-6}. First, based on \eqref{eq:UCCF-6}, an
autocorrelation matrix of $\pmb{Y}_m$ can be obtained as
\begin{align}\label{eq:UCCF-10}
\pmb{R}_{m}=\pmb{Y}_m\pmb{Y}_m^H/{\tau_p}
\end{align}
Note that, if some observations for data transmission following channel training are available, the estimation of $\pmb{R}_{m}$ can be enhanced by invoking
the data-related observations in \eqref{eq:UCCF-10}. Then, using $\pmb{R}_m$, the matrix
in the bracket of \eqref{eq:UCCF-9} is replaced by
$(\pmb{I}_{\tau_p}\otimes \pmb{R}_{m})$, where $\otimes$ represents
the Kronecker operation.

After the channels of UEs are respectively estimated at APs, APs can
forward the estimated channels to CPU, if required. Alternatively, APs
may forward the observations seen in \eqref{eq:UCCF-6} to CPU, and CPU
carries out the channel estimation. In this case, the observations
obtained by CPU may be expressed as
\begin{align}\label{eq:UCCF-11}
\pmb{Y}'_{m}=a_{cm}\pmb{Y}_m,~m\in\mathcal{M}
\end{align}
where $a_{cm}$ is the channel gain between CPU and AP $m$. In ideal
case, or when CPU and AP $m$ have the ideal knowledge about $a_{cm}$,
$a_{cm}=1$ can be assumed. After CPU obtains $\pmb{Y}'_{m}$, it can
estimate the channels of the UEs associated with AP $m$, in the same
way as done in \eqref{eq:UCCF-8} or \eqref{eq:UCCF-9}.

After obtaining the estimate $\hat{\pmb{h}}_{mk}$ to the CIR of
${\pmb{h}}_{mk}$, the channel gains of $N$ subcarriers can be obtained
as
\begin{align}\label{eq:UCCF-12}
\hat{\pmb{h}}_{f,mk}=\pmb{F}_N\pmb{\Psi}_{mk}\hat{\pmb{h}}_{mk},~m\in\mathcal{M}; k\in\mathcal{K}_m
\end{align}
which can be applied for UL detection and DL transmission.

It is worth noting that in
$\hat{\pmb{h}}_{f,mk}=\sqrt{\hat{g}_{mk}}\hat{\tilde{\pmb{h}}}_{mk}$,
the large-scale propagation pathloss and shadowing, i.e.,
$\hat{g}_{mk}$, is the same for all subcarriers, but the small-scale
fading gains of different subcarriers in $\hat{\tilde{\pmb{h}}}_{mk}$
may be different.

\index{Channel!estimation|)}
\section{Access-Point Association of User Equipments}\label{section-6G-4.4}\index{Association!user equipment|(}

Optimal association of UEs with APs with the objective to, e.g.,
maximize system sum-rate, energy efficiency, etc., under various
constraints, e.g., individual UEs' minimum rate and power, total AP
transmit power, backhaul resources, etc., can be extremely
complicated, if the number of APs and the number of UEs are relatively
big. The complexity roots in that the optimization of UE association
is often coupled with the other optimizations at system and link
levels. Hence, in practice, simple low-complexity association methods
are desired, and to achieve this, the UE association process is
usually separated from the processes of system and link
optimization. Below are some low-complexity UE association methods.

The first one is the simplest distance-based UE association. With this
method, a distance $R_{th}$ is initialized for the association
decision making. A UE $k$ is associated with all the APs that have
their distances from UE $k$ not exceeding $R_{th}$, which form a set
$\mathcal{M}_k$. If there is no AP satisfying the condition, UE $k$ is
then associated with an AP that has the minimum distance from the UE,
provided that the minimum QoS required by UE $k$ can be met via, such
as, power-control or/and variable data rate transmission, etc.,
techniques. Otherwise, if the minimum QoS requirement of UE $k$ is
unable to be guaranteed in any way, the UE will not be associated with
any APs, and may be disconnected from the network.

The second method carries out UE association based on the large-scale
fading, i.e., based on $g_{mk}$ seen in \eqref{eq:UCCF-1}. To
implement this method, for each UE $k$, $\{g_{mk}\}$ are first listed
in descending order. Then, the APs are selected to be included in
$\mathcal{M}_k$ according to the list, in the order from the best
(largest $g_{mk}$) to worst (smallest $g_{mk}$), until the stop
conditions are met. The stop conditions may be the number of APs,
$g_{mk}$ is lower than a pre-set threshold, etc.


In some UCCF networks, APs are supposed to be densely deployed. This
kind of networks have the feasibility for the implementation of
wireless sensing\index{Wireless sensing}, which can benefit from the
relatively easy access of LoS signals and the availability of many
reference points (such as APs) for variable sensing. Furthermore, with
the aid of integrated sensing and communications
(ISAC)\index{Integrated sensing and communications (ISAC)}, it is
achievable that APs and UEs are able to know the detailed
communication environments, in addition to the accurate positions of
them. With this positioning and environmental information, it is
possible for a UE to find the best physical paths for it to propagate
signals to APs. Similarly, APs can use the best paths to send
information to UEs. Hence, with the aid of wireless sensing or ISAC,
UEs can collaborate with APs to settle down their associations based
on the sensed information about communication environments as well as
the APs and UEs themselves.  \index{Association!user equipment|)}
\section{Uplink Detection and Optimization}\label{section-6G-4.5}

This section first provides the principles of UL detection in UCCF
systems via analyzing several detection methods. Then, the UL resource
optimization is discussed.

\subsection{Uplink Detection Schemes}\label{section-6G-4.5.1}

To study the UL detection and optimization, assume that UE $k$ is
associated with the APs in $\mathcal{M}_k$, which includes AP $m$. Let
the frequency-domain symbol vector sent by UE $k$ on $N_k$ subcarriers
is denoted as $\pmb{x}_k\in\mathcal{C}^{N_k\times 1}$, which satisfies
$E[\pmb{x}_k\pmb{x}_k^H]=\pmb{I}_{N_k}$. Assume that the maximum
transmit power of UEs is $P_u$, and the $k$th UE's transmit power is
$P_k=\eta_kP_u$, where $\eta_k\leq 1$ is the power-control
coefficient.\index{Power-control!coefficient} Then, by following the
principles of OFDM in Chapter~{\bf (OFDM)}, the corresponding
observations obtained by AP $m$ from $N$ subcarriers can be
represented as
\begin{align}\label{eq:UCCF-13}
\pmb{y}_{m}=\sum_{k\in\mathcal{K}}\pmb{H}_{mk}\pmb{\Phi}_{k}\pmb{\eta}_k^{1/2}\pmb{x}_k+\pmb{n}_m,~m=1,2,\ldots,M
\end{align}
where
$\pmb{H}_{mk}=\textrm{diag}(\pmb{F}_N\pmb{\Psi}_{mk}\pmb{h}_{mk})$,
with the diagonal elements denoting the channel gains of corresponding
subcarriers, and $\pmb{h}_{mk}$ is given by \eqref{eq:UCCF-1},
including both the large-scale propagation pathloss and shadowing, as
well as the small-scale fading gains. It can be understood that the
large-scale propagation pathloss and shadowing are UE dependent but
not subcarrier dependent, while the small-scale fading gains of
different subcarriers of a UE may be different. Again,
$\pmb{\Phi}_{k}$ executes subcarrier-allocation to choose $N_k$
subcarriers for transmitting the information of UE
$k$. $\pmb{\eta}_k=\textrm{diag}\{\eta_{k1},\eta_{k2},\cdots,\eta_{kN_k}\}$,
satisfying $\sum_{n=1}^{N_k}=\eta_k$, controls the power assigned to
the $N_k$ subcarriers.  $\pmb{n}_m$ is the Gaussian noise distributed
with mean zero and a covariance matrix of $\gamma_u^{-1}\pmb{I}_N$,
where $\gamma_u=P_u/\sigma^2$.

Below the principles of three detection schemes are analyzed,
including the global MMSE (GMMSE) detection, local MMSE (LMMSE)
detection and the AP message-passing (APMP) detection. To illustrate
the principles, channels are assumed to be ideally estimated, backhaul
links are ideal for information exchange between CPU and APs, and the
UCCF network is operated in the ideal synchronization state.

\subsubsection{Global MMSE Detection}\label{section-6G-4.5.1.1}\index{MMSE detection!global|(}

GMMSE detector is operated at CPU, which uses all the observations
from $M$ APs to detect any a UE's information. In other words, APs do
not process their observations locally, but directly forward
$\{\pmb{y}_m\}$ in \eqref{eq:UCCF-13} to CPU. At CPU, let
$\pmb{y}=[\pmb{y}_1^T,\pmb{y}_2^T,\ldots,\pmb{y}_M^T]^T$,
$\pmb{n}=[\pmb{n}_1^T,\pmb{n}_2^T,\ldots,\pmb{n}_M^T]^T$,
$\pmb{H}_k=\left[\pmb{H}_{1k}^T,\pmb{H}_{2k}^T,\ldots,\pmb{H}_{Mk}^T\right]^T$. Then,
$\pmb{y}$ can be written as
\begin{align}\label{eq:UCCF-14}
\pmb{y}=\sum_{k\in\mathcal{K}}\pmb{H}_k\pmb{\Phi}_{k}\pmb{\eta}_k^{1/2}\pmb{x}_k+\pmb{n}
\end{align}    

To detect the information of UE $k$, CPU forms the decision variable
in MMSE principle as
\begin{align}\label{eq:UCCF-15}
\hat{\pmb{x}}_k=\pmb{W}_k^H\pmb{y},~k\in\mathcal{K}
\end{align}
where $\pmb{W}_k$ can be expressed as~\cite{Lie-Liang-MC-CDMA-book}
\begin{align}\label{eq:UCCF-16}
\pmb{W}_k=\pmb{R}_y^{-1}\pmb{R}_{yk}
\end{align}
with $\pmb{R}_y$ being autocorrelation matrix\index{Autocorrelation}
of $\pmb{y}$, given by
\begin{align}\label{eq:UCCF-17}
\pmb{R}_y=E\left[\pmb{y}\pmb{y}^H\right]=\sum_{k\in\mathcal{K}}\pmb{H}_k\pmb{\Phi}_{k}\pmb{\eta}_k\pmb{\Phi}_{k}^T\pmb{H}_k^H+\gamma_u^{-1}\pmb{I}_{MN}
\end{align}
and $\pmb{R}_{yk}$ being the cross-correlation
matrix\index{Cross-correlation} between $\pmb{y}$ and $\pmb{x}_k$,
having the expression of
\begin{align}\label{eq:UCCF-18}
\pmb{R}_{yk}=E\left[\pmb{y}\pmb{x}_k^H\right]=\pmb{H}_k\pmb{\Phi}_{k}\pmb{\eta}_k^{1/2}
\end{align}
Hence, $\pmb{W}_k$ is
\begin{align}\label{eq:UCCF-19}
\pmb{W}_k=\left(\sum_{l\in\mathcal{K}}\pmb{H}_l\pmb{\Phi}_{l}\pmb{\eta}_l\pmb{\Phi}_{l}^T\pmb{H}_l^H+\gamma_u^{-1}\pmb{I}_{MN}\right)^{-1}\pmb{H}_k\pmb{\Phi}_{k}\pmb{\eta}_k^{1/2}
\end{align}

Note that, in addition to mitigating MUI, the GMMSE detector described
in \eqref{eq:UCCF-15}-\eqref{eq:UCCF-19} is also able to suppress the
embedded inter-carrier interference (ICI)\index{inter-carrier
  interference (ICI)}, if it exists. If subcarriers are ideally
orthogonal, the GMMSE detector can be separated into $N$ GMMSE
detectors operated in parallel, each is for detecting the symbols
conveyed on one subcarrier. It can also be shown that the complexity
required by the $N$ separate GMMSE detectors is much lower than that
of the joint GMMSE detector. Moreover, corresponding to one
subcarrier, such as, $n$, the formulas for the GMMSE detector are
similar as that provided in \eqref{eq:UCCF-14}-\eqref{eq:UCCF-19},
only with \eqref{eq:UCCF-13} changed to
\begin{align}\label{eq:UCCF-20}
y_{mn}=\sum_{k\in\mathcal{K}}\sqrt{\eta_{kn}}h_{mn,k}\delta_{kn}x_{kn}+{n}_{mn},n\in\mathcal{N};~m\in\mathcal{M}
\end{align}
where $h_{mn,k}$ is the channel gain of the $n$th subcarrier of UE
$k$, $x_{kn}$ is the symbol sent on subcarrier $n$ by UE $k$,
$\delta_{kn}=1$ if UE $k$ is assigned subcarrier $n$, otherwise,
$\delta_{kn}=0$. 

Expressing $\pmb{W}_k=\left[\pmb{w}_{k1},\pmb{w}_{k2},\ldots,\pmb{w}_{kN_k}\right]$. It can be shown that $\pmb{w}_{ki}$ for detecting the $i$th symbol of UE $k$ is 
\begin{align}\label{eq:UCCF-21}
\pmb{w}_{ki}=\sqrt{{\eta}_{ki}}\left(\sum_{l\in\mathcal{K}}\pmb{H}_l\pmb{\Phi}_{l}\pmb{\eta}_l\pmb{\Phi}_{l}^T\pmb{H}_l^H+\gamma_u^{-1}\pmb{I}_{MN}\right)^{-1}\pmb{H}_k\pmb{\phi}_{ki}
\end{align}
where $\pmb{\phi}_{ki}$ is the $i$th column of $\pmb{\Phi}_{k}$. The SINR is~\cite{Lie-Liang-MC-CDMA-book} 
\begin{align}\label{eq:UCCF-22}
\gamma_{ki}&\left(\{\pmb{\eta}_l\},\{\pmb{\Phi}_l\}\right)={\eta_{ki}}\pmb{\phi}_{ki}^H\pmb{H}_k^H\pmb{R}^{-1}_{ki}\pmb{H}_k\pmb{\phi}_{ki}\nonumber\\
&={\eta_{ki}}\pmb{\phi}_{ki}^H\pmb{H}_k^H\left(\sum_{l\in\mathcal{K}}\pmb{H}_l\pmb{\Phi}_{l}\pmb{\eta}_l\pmb{\Phi}_{l}^T\pmb{H}_l^H-\eta_{ki}\pmb{H}_k\pmb{\phi}_{ki}\pmb{\phi}_{ki}^T\pmb{H}_k^H+\gamma_u^{-1}\pmb{I}_{MN}\right)^{-1}\pmb{H}_k\pmb{\phi}_{ki}
\end{align}
where
$\pmb{R}_{ki}=\sum_{l\in\mathcal{K}}\pmb{H}_l\pmb{\Phi}_{l}\pmb{\eta}_l\pmb{\Phi}_{l}^T\pmb{H}_l^H-\eta_{ki}\pmb{H}_k\pmb{\phi}_{ki}\pmb{\phi}_{ki}^T\pmb{H}_k^H+\gamma_u^{-1}\pmb{I}_{MN}$
is the autocorrelation matrix of interference plus noise when
detecting symbol $i$ of UE $k$. Accordingly, the
sum-rate\index{Sum-rate} of UCCF system is
\begin{align}\label{eq:UCCF-23}
R\left(\{\pmb{\eta}_l\},\{\pmb{\Phi}_l\}\right)=\sum_{k\in\mathcal{K}}\sum_{i=1}^{N_k}\log_2\left[1+\gamma_{ki}\left(\{\pmb{\eta}_l\},\{\pmb{\Phi}_l\}\right)\right]
\end{align}
which is a function of the power-control coefficients in
$\{\pmb{\eta}_l\}$, and the subcarrier mapping matrices
$\{\pmb{\Phi}_l\}$ that implement subcarrier-allocation.

GMMSE detector has the potential to achieve promising performance in
the case that $M\geq K$, and the UEs are evenly distributed in the
network. However, it is hard to implement. As shown in $\pmb{H}_k$
defined above \eqref{eq:UCCF-14}, it includes not only the channels
between UE $k$ and the APs in $\mathcal{M}_k$, i.e., the APs that UE
$k$ is associated with, but also the channels in
$\bar{\mathcal{M}}_k$, which represents the APs that UE $k$ is not
associated with. In practice, the channels between UE $k$ and the APs
in $\bar{\mathcal{M}}_k$ are supposed to be weak. Hence, these
channels can hardly be estimated with sufficient accuracy for
information detection. Furthermore, GMMSE detector needs to invert a
matrix of $(MN\times MN)$-dimensional, or $N$ matrices of $(M\times
M)$-dimensional in the case of using $N$ parallel detectors, with each
for one subcarrier, the detection complexity may be extreme for
practical implementation.  To circumvent these problems, below the
local detection schemes that are UEs' location-aware are
considered. Here, the local detection of UE $k$ means that its
information is detected only based on the observations collected from
its associated APs.

\index{MMSE detection!global|)}

\subsubsection{Local MMSE Detection}\label{section-6G-4.5.1.2}\index{MMSE detection!local|(}

LMMSE detection can be operated at CPU or at
APs~\cite{9220147}. First, when operated at CPU, let us express the
inverted matrix $\pmb{R}_y^{-1}$ in \eqref{eq:UCCF-19} as
$\pmb{R}_y^{-1}=\left[\pmb{Q}_1,\pmb{Q}_2,\ldots,\pmb{Q}_M\right]$,
where $\pmb{Q}_m$ are $(MN\times N)$ matrices. Then, when only
considering the APs in $\mathcal{M}_k$ to detect UE $k$, the weight
matrix can be approximated by
\begin{align}\label{eq:UCCF-24}
\pmb{W}_k\approx\sum_{m\in\mathcal{M}_k}\pmb{Q}_m\pmb{H}_{mk}\pmb{\Phi}_{k}\pmb{\eta}^{1/2}
\end{align}

Note that the autocorrelation matrix $\pmb{R}_y$ in \eqref{eq:UCCF-19}
can be kept the same for the LMMSE detection, as it can be directly
estimated from the observations provided by APs as
\begin{align}\label{eq:UCCF-25}
\pmb{R}_y=\frac{1}{U}\sum_{u=1}^U\pmb{y}(u)\pmb{y}^H(u)
\end{align}
where $\{\pmb{y}(u)\}$ are observations of $\pmb{y}$ obtained in $U$ OFDM symbol periods. The estimate to $\pmb{R}_y$ becomes more accurate, as $U$ increases.

Although the approximated weight matrix in \eqref{eq:UCCF-24} does not
use all the columns of $\pmb{R}_y^{-1}$, but each $\pmb{Q}_m$ has the
corresponding elements in the $MN$ rows of $\pmb{R}_y^{-1}$. Hence,
the LMMSE detector with the $\pmb{W}_k$ of \eqref{eq:UCCF-24} is
capable of effectively suppressing MUI. In comparison with the ideal
GMMSE detector in Section~\ref{section-6G-4.5.1.1}, the performance
loss of the LMMSE detector lies in the fact that some APs may receive
certain power from UE $k$, but they are not included in
$\mathcal{M}_k$, and hence the power is not exploited for the
detection of UE $k$. Therefore, in the UE association stage, the APs
making noticeable contribution to UE $k$'s detection performance are
expected to be included in $\mathcal{M}_k$, if the complexity for,
such as, channel estimation, is allowable.

The LMMSE detector for UE $k$ may be implemented at CPU by only
considering the observations from the APs that UE $k$ is associated
with. Alternatively, it can be implemented at an AP (e.g., the best
one in terms of UE $k$) via AP cooperation, which is supported by
information exchange between APs (usually between the geographically
adjacent APs). Assume that the information exchange between APs and
CPU, or between APs is ideal. Let, for the $|\mathcal{M}_k|$ APs
associated by UE $k$, define
$\pmb{y}=\left[\pmb{y}_1^T,\pmb{y}_2^T,\ldots,\pmb{y}_{|\mathcal{M}_k|}^T\right]^T$,
$\pmb{n}=\left[\pmb{n}_1^T,\pmb{n}_2^T,\ldots,\pmb{n}_{|\mathcal{M}_k|}^T\right]^T$
and
$\pmb{H}_k=\left[\pmb{H}_{1k}^T,\pmb{H}_{2k}^T,\ldots,\pmb{H}_{|\mathcal{M}_k|k}^T\right]^T$. Then,
we have a same representation as \eqref{eq:UCCF-14} for carrying out
the LMMSE detection. Furthermore, the LMMSE detector has the same set
of formulas as shown in \eqref{eq:UCCF-15}-\eqref{eq:UCCF-23}, but the
dimension $MN$ of the matrices invoked in the GMMSE detector is
reduced to $|\mathcal{M}_k|N$ in the LMMSE detector. Since
$|\mathcal{M}_k|<<M$, it can be expected that the LMMSE detector
performs worse than the ideal GMMSE detector, due to its deteriorated
MUI suppressing capability as compared to the ideal GMMSE
detector. However, this LMMSE detector has significantly lower
complexity than the GMMSE detector and the LMMSE detector using the
weighting of \eqref{eq:UCCF-24}.

In a LMMSE detection scenario where main detection processing is
carried out at APs, if APs do not cooperate with each other, the MMSE
processing has to be built on $\pmb{y}_m$ of \eqref{eq:UCCF-13}. In
this case, the $m$th AP can first form a local estimator corresponding
to UE $k$ as
\begin{align}\label{eq:UCCF-26}
\pmb{z}_{mk}=\pmb{W}_{mk}^H\pmb{y}_m,~m\in\mathcal{M}_k,~k\in\mathcal{K} 
\end{align}   
where the weight matrix $\pmb{W}_{mk}$ can be derived based on \eqref{eq:UCCF-13}, and is given by
\begin{align}\label{eq:UCCF-27}
\pmb{W}_{mk}=\left(\sum_{l\in\mathcal{K}}\pmb{H}_{ml}\pmb{\Phi}_{l}\pmb{\eta}_l\pmb{\Phi}_{l}^T\pmb{H}_{ml}^H+\gamma_u^{-1}\pmb{I}_{N}\right)^{-1}\pmb{H}_{mk}\pmb{\Phi}_{k}\pmb{\eta}_k^{1/2}
\end{align}
Then, $\pmb{z}_{mk}$ is forwarded to CPU, where a final decision variable vector for UE $k$'s $N_k$ symbols can be formed as
\begin{align}\label{eq:UCCF-28}
\pmb{z}_k=\sum_{m\in\mathcal{M}_k}\pmb{\Lambda}_{mk}\pmb{z}_{mk},~k\in\mathcal{K} 
\end{align}   
where $\pmb{\Lambda}_{mk}=\textrm{diag}\{\lambda_{mk,1},\lambda_{mk,2},\ldots,\lambda_{mk,N_k}\}$ are the weights used by CPU to combine $\pmb{z}_{mk}$.

Depending on the channel knowledge available to CPU, the combining
weights in \eqref{eq:UCCF-28} can be set in different ways. First,
when CPU has no knowledge at all, it chooses
$\pmb{\Lambda}_{mk}=\pmb{I}_{N_k}/|\mathcal{M}_k|$. Second, express
$\pmb{z}_{mk}$ in \eqref{eq:UCCF-26} as
$\pmb{z}_{mk}=\pmb{A}_{mk}\pmb{x}_k+\pmb{n}_{I,mk}+\pmb{n}_{mk}$,
where $\pmb{n}_{I,mk}$ is interference. Then, when CPU knows
$\pmb{A}_{mk}$ and the covariance matrix of interference plus noise,
which is expressed as $\pmb{C}_{mk}$, it can set
\begin{align}\label{eq:UCCF-29}
\pmb{\Lambda}_{mk}=\pmb{C}_{mk}^{-1}\pmb{A}_{mk}^H/|\mathcal{M}_k|
\end{align}
to implement maximal ratio combining (MRC)\index{Maximal ratio
  combining (MRC)}, which maximize SINR~\cite{Proakis-5th}. To
compute \eqref{eq:UCCF-29}, the knowledge of both the large-scale and
small-scale fading are required.  Third, if CPU only has the knowledge
of the large-scale fading, it may combine $\pmb{z}_{mk}$ as
\begin{align}\label{eq:UCCF-30}
\pmb{z}_k=\sum_{m\in\mathcal{M}_k}\frac{g_{mk}}{\sum_{l\in\mathcal{M}_k}g_{lk}}\pmb{z}_{mk}~~\textrm{or}~~\pmb{z}_k=\sum_{m\in\mathcal{M}_k}\frac{\sqrt{g_{mk}}}{\sum_{l\in\mathcal{M}_k}\sqrt{g_{lk}}}\pmb{z}_{mk}
\end{align}   
where $g_{lk}$ is defined with \eqref{eq:UCCF-1}.

\index{MMSE detection!local|)}
\subsubsection{Access Point Message Passing Detection}\label{section-6G-4.5.1.3}\index{Detection!access point message passing|(}

%
\begin{figure}[th]
  \begin{center} 
   \includegraphics[width=0.65\linewidth]{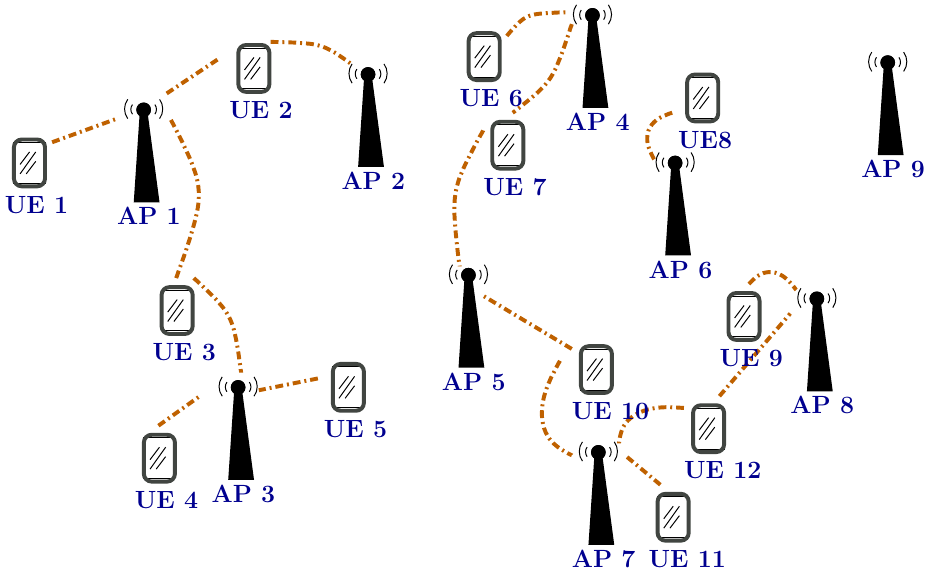}
   \end{center} 
  \caption{An example of UCCF network to illustrate the application of
    the message-passing algorithm (MPA)\index{Message-passing
      algorithm (MPA)} for UL detection.}
  \label{figure-UCCF-MPA-factor-graph}
\end{figure}
\begin{figure}[th]
  \begin{center} 
   \includegraphics[width=0.95\linewidth]{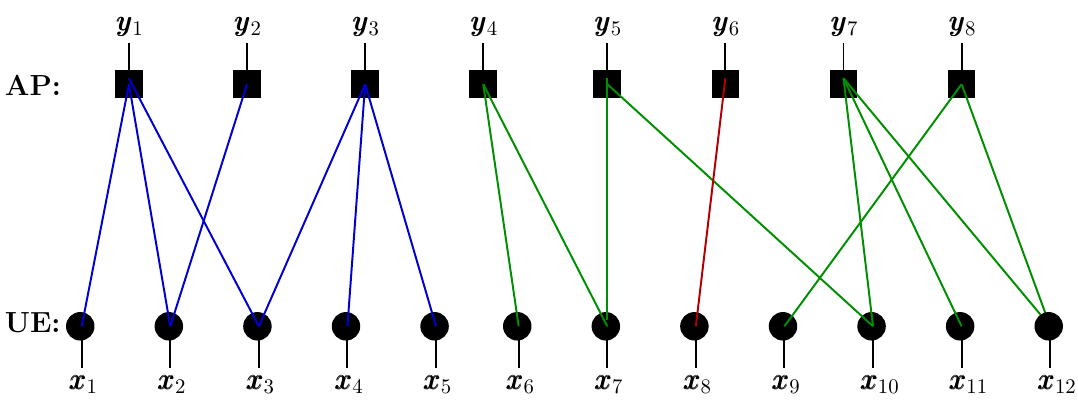}
   \end{center} 
  \caption{Factor graph\index{Factor graph} showing the relationship between APs and UEs in the UCCF network of Fig.~\ref{figure-UCCF-MPA-factor-graph}.}
  \label{figure-UCCF-MPA-factor-graph-1}
\end{figure}

In a densely deployed UCCF network, each AP typically serves a small
number UEs, and each UE is only associated with a small number of APs
close to the UE. Hence, when abstracting APs as function nodes (FNs)
and UEs as variable nodes (VNs), a sparse factor
graph~\cite{Shu_Lin-Ryan} \index{Factor graph!sparse}can be
constructed to explain the relationship of APs and UEs in the UCCF
network. For example, Fig.~\ref{figure-UCCF-MPA-factor-graph} shows a
UCCF network with 9 APs and 12 UEs, as well as the association states
of the UEs. Accordingly, the factor graph describing the relationship
between APs/UEs is shown in Fig.~\ref{figure-UCCF-MPA-factor-graph-1},
where AP $i$ and UE $j$ are connected by an edge, if UE $j$ is
associated with AP $i$. Based on
Fig.~\ref{figure-UCCF-MPA-factor-graph} and
Fig.~\ref{figure-UCCF-MPA-factor-graph-1}, the factor graph describing
a UCCF network has the following characteristics:
\begin{itemize}

\item The factor graph is sparse.

\item In comparison with the well-designed regular factor graphs
  describing low-density parity check (LDPC)
  codes~\cite{Shu_Lin-Ryan}, the edges connecting FNs and VNs are
  irregular, which are depended on the distributions of UEs and APs,
  as well as the association rules applied.

\item The factor graph may be divided into several sub-graphs that are
  independent of each other. Seen in
  Fig.~\ref{figure-UCCF-MPA-factor-graph-1}, there are three
  sub-graphs, having the FNs $\{1,2,3\}$, $\{4,5,7,8\}$ and $\{6\}$,
  respectively. This reflects that the UEs in different sub-region are
  associated with the isolated groups of APs, due to their
  geographical separation. Hence, the interference between two UEs
  belonging to two sub-graphs is low or no interference between them.

\item It is possible that one sub-graph contains only one UE connected
  to one AP, e.g., UE 8 to AP 6 in
  Fig.~\ref{figure-UCCF-MPA-factor-graph-1}, or to multiple
  APs. Furthermore, it is possible that an AP has no UEs associated
  with, e.g., AP 9 in Fig.~\ref{figure-UCCF-MPA-factor-graph}. Hence,
  the corresponding FN has no edges and in the case, the AP can be
  switched off to save energy.

\end{itemize}

The access point message passing (APMP) detector~\cite{9220147}
exploits the principles of message passing algorithm (MPA)\index{
  Message passing algorithm (MPA)} for signal detection. It is
operated on the factor graph, as shown in
Fig.~\ref{figure-UCCF-MPA-factor-graph-1} for example. In detail, the
APMP detection process is described as follows, where intrinsic and
extrinsic information can be calculated in the principles, for
example, as detailed
in~\cite{Shu_Lin-Ryan,book:xiaodong-wang,9220147,8166802}.\index{Intrinsic
  information}\index{Extrinsic information} To explain, let us assume
an example that APs $i$ and $j$ are two adjacent APs, and UE $k$ has
been associated with both of them.
\begin{enumerate}

\item After UE association, APs report to CPU their associated UEs,
  i.e., send $\mathcal{K}_m, \forall m\in\mathcal{M}$, to CPU.

\item Based on $\{\mathcal{K}_m\}$, CPU builds the factor graph and
  informs each AP its adjacent APs. Two APs are defined as adjacent
  APs, if they monitor one to several common UEs. Hence, in the
  example, APs $i$ and $j$ are adjacent APs. Later during detection,
  two adjacent APs will exchange the information about their commonly
  monitored UEs to improve the detection reliabilities of these UEs.

\item Initially, each of APs calculates the intrinsic information of
  the symbols (or bits) of the UEs associated with it, based on its
  locally received observations.

\item Adjacent APs exchange the information of their commonly
  monitored UEs. Considering the example, APs $i$ and $j$ exchange
  their information about UE $k$. AP $i$ and AP $j$ may also receive
  information about UE $k$ from their other adjacent APs.

\item After receiving the information from adjacent APs, for each of
  its associated UEs, an AP computes, respectively, the total
  information of the individual symbols (or bits) of the UE.

\item Then, each AP feeds back the extrinsic information of symbols (bits) of an UE to its adjacent APs that also monitor the UE. Here, the extrinsic information of a symbol (or bit) passed to a specific AP is equal to the difference between the total information of the symbol (or bit) and the information of that symbol (or bit) previously received from the same AP. 

Considering AP $i$ in the example, during an iteration, it first calculates the total information of a symbol (or bit) based on its intrinsic information and the extrinsic information received from other APs. Then, the extrinsic information fed back to AP $j$ is the difference between the total information of the symbol (or bit) and the information about the symbol (or bit) that AP $i$ previously received from AP $j$.

\item The iteration process 4) - 6) is repeated until no increase of information is available, or the allowed number of iterations is reached.   

\item Finally, the symbols (or bits) of a specific UE are decided by one of its APs.    

\end{enumerate}

From the above description, the APMP detector is mainly operated by
APs, CPU is only required to build and maintain the factor graph, and
update it to APs when there are changes occurred in the network. APs
are only required to exchange information with their adjacent
APs. Hence, it is expected that the demand on backhaul resources is
moderate. In addition, it is well-known, also evidenced by the results
in literature, such as references
\cite{Shu_Lin-Ryan,book:xiaodong-wang,9220147,8166802}, that the
performance achieved by the MPA-assisted detection is near-optimum,
close to that achieved by the optimum maximum likelihood detection.

\index{Detection!access point message passing|)}
\subsection{Uplink Resource Optimization}\label{section-6G-4.5.2}

Uplink resource optimization in the considered OFDM signaling UCCF
system includes UE association, subcarrier-allocation and
power-control. Optimization constraints include the available transmit
power of UEs and, possibly, some service requirements, such as,
minimum rate, delay, etc., of individual UEs. Hence, the optimum
solutions to maximize UL sum-rate are required to solve, for example,
the problem:\index{Optimization problem!maximize sum-rate}
\begin{subequations}\label{eq:UCCF-RO31}
\begin{align}
\{\zeta_{mk}^*\},&\{\delta_{nk}^*\},\{\eta_{kn}^*\}\nonumber\\
&=\arg\max_{\{\delta_{kn}\},\{\zeta_{mk}\},\{\eta_{kn}\}}\left\{\sum_{k\in\mathcal{K}}\sum_{n\in\mathcal{N}}\delta_{kn}\log_{2}\left(1+\gamma_{kn}\left(\{\zeta_{mk}\},\{\eta_{kn}\}\right)\right)\right\}\\
s.t.~~& \zeta_{mk}\in\{0,1\},~\forall m\in\mathcal{M}, k\in\mathcal{K}\\
& \delta_{kn}\in\{0,1\},~\forall k\in\mathcal{K}, n\in\mathcal{N}\\
&\sum_{n\in\mathcal{N}}\eta_{kn}=\eta_k\leq 1,~\forall k\in\mathcal{K}\\
&\sum_{n\in\mathcal{N}}\delta_{kn}\log_{2}\left(1+\gamma_{kn}\left(\{\zeta_{mk}\},\{\eta_{kn}\}\right)\right)\geq R_k^{\min},~\forall k\in\mathcal{K}
\end{align}
\end{subequations}
In this optimization problem, \eqref{eq:UCCF-RO31}(b) controls UE
association, \eqref{eq:UCCF-RO31}(c) carries out
subcarrier-allocation, \eqref{eq:UCCF-RO31}(d) constrains the power
assigned to the active subcarriers of UEs, and finally,
\eqref{eq:UCCF-RO31}(e) guarantees that each UE is at least provided
with the minimum service rate as required. In the above formula,
$\gamma_{kn}\left(\{\zeta_{mk}\},\{\eta_{kn}\}\right)$ is the SINR of
subcarrier $n$ of UE $k$, which is dependent on the detection method
employed, it is a function of the APs that UE $k$ is associated with,
and the power allocated to the subcarriers assigned to UE
$k$. Finally, $\{\zeta_{mk}^*\},~\{\delta_{nk}^*\}$ and
$\{\eta_{kn}^*\}$ are the sets of solutions obtained after the
optimization.

Due to the involved binary optimization from UE association and
subcarrier-allocation, the problem of \eqref{eq:UCCF-RO31} is NP hard
to solve. Hence, low-complexity and yet efficient methods are aimed
at. The simplest method might be optimizing UE association,
subcarrier-allocation and power-allocation in succession
separately. Specifically, UE association can be executed first, for
example, by applying a method discussed in
Section~\ref{section-6G-4.4}. After UE association, referring to
Chapters~{\bf (Subcarrier-Allocation)}, subcarrier-allocation can then
be carried out by considering the rate requirements of individual UEs,
the relative distributions of APs and UEs, possible interfernece
between UEs, etc. Finally, power is allocated to the assigned
subcarriers of a UE to maximize the total rate of the UE.

Instead of maximizing sum-rate, energy-efficiency can be an
alternative utility function to maximize in Problem
\eqref{eq:UCCF-RO31}. In the UCCF systems as considered, the power
consumption includes the circuit power by UEs, APs and CPU, the
transmit power between UEs and APs, that between APs and CPU, and
possibly, the circuit/transmit power for AP cooperation. Due to the
fact that sum-rate is a function of power, the energy-efficiency is a
fractional objective function, which is non-concave and hard to solve
using standard programming methods, even only power-control is
considered~\cite{8676377}. Hence, the optimization problem has to be
divided into a range of sub-problems to be tackled, such as, by the
means of the sequential optimization\index{Optimization!sequential}
operated on the lower-bound of the energy-efficiency
function~\cite{8676377}. Note that, maximizing the energy-efficiency
of UCCF networks may result in that some APs are switched off to save
energy.

Another optimization objective may be to maximize the minimum rate or
minimum SINR of UEs, forming the max-min problem\index{Optimization
  problem!max-min}, as shown in Chapter~{\bf
  (Power-Allocation)}. Again, the join optimization of UE association,
subcarrier-allocation and power-control is extremely difficult to
achieve. They may instead be optimized successively, following the
order of UE association, subcarrier-allocation and power-control. In
Chapter~{\bf (Power-Allocation)}, it is shown that the max-min
optimization makes a trade-off on efficiency for fairness. However, in
UCCF networks, each UE can be expected to be located close to one and
even several APs. Hence, every UE is a strong UE with regard to some
APs, resulting in that the greedy-based subcarrier-allocation should
be near-optimum. Consequently, each UE is able to send information
over high-quality channels. In this case, the followed
power-allocation is relatively simple to implement, and the
power-control optimized in max-min principle can also be near-optimum
in the sense of maximal sum-rate or maximal energy efficiency, in
addition to the fairness provided by the max-min optimization. In
plain language, a UCCF network is capable of enabling the system-level
efficiency and the individual-level fairness to be simultaneously
near-optimum.







\section{Downlink Transmission and Optimization}\label{section-6G-4.6}

Considering the transmitter preprocessing\index{Transmitter
  preprocessing} in MMSE principle (TMMSE), in this section, the
principles of DL transmission and optimization are considered. For
clarity and generality, subcarrier level and OFDM symbol level
processings are analyzed in parallel.\index{Transmitter
  preprocessing!subcarrier level}\index{Transmitter preprocessing!OFDM
  symbol level} Note that, when only one subcarrier is invoked, the
model and analysis become the same of a DL space division multiplexing
(SDM) model\index{Space division multiplexing (SDM)}. By contrast, if
OFDM is invoked, the model implements both SDM and frequency-division
multiplexing (FDM)\index{Frequency-division multiplexing (FDM)}. Note
that, if an OFDM system exists ICI, the preprocessing at OFDM symbol
level is required to suppress this interference.

\subsection{Downlink Transmission with Precoding}\label{section-6G-4.6.1}\index{Downlink!precoding|(}

Following the assumptions and settings applied in the UL analysis, let
us assume that a subcarrier symbol or an OFDM symbol to be sent to UE
$k$ by AP $m$ is expressed as $x_{k}\in\mathcal{C}$ or
$\pmb{x}_{k}\in\mathcal{C}^{N_k\times 1}$, which satisfies
$E[|x_k|^2]=1$ or $E[\|\pmb{x}_k\|^2]=N_k$. Assume that the design is
implemented by CPU and, for the moment, assume that CPU has the ideal
channel knowledge from any of APs to any of UEs. In other words, all
$M$ APs are assumed to simultaneously transmit signals to all $K$ UEs.
Then, after preprocessing the data symbols of the $K$ UEs, the signal
sent by AP $m$, $m\in\mathcal{M}$, can be expressed as
\begin{subequations}\label{eq:UCCF-31}
\begin{align}
{s}_m=&A_0\sum_{k\in\mathcal{K}}p_{mk}x_k\\
\tilde{\pmb{s}}_m=&A_0\sum_{k\in\mathcal{K}}\pmb{P}_{mk}\pmb{\Phi}_k\pmb{x}_k
\end{align}  
\end{subequations}
where $\pmb{a}$ and $\tilde{\pmb{a}}$ are used to indicate the
subcarrier level and OFDM symbol level, respectively. In
\eqref{eq:UCCF-31}, $s_m\in\mathcal{C}$ and
$\tilde{\pmb{s}}_m\in\mathcal{C}^{N\times 1}$, $p_{mk}\in\mathcal{C}$
and $\pmb{P}_{mk}=\textrm{diag}\{\pmb{p}_{mk}\}\in\mathcal{C}^{N\times
  N}$ are preprocessing scalar and matrices. Again, the mapping
matrices $\{\pmb{\Phi}_k\}$ implement DL subcarrier-allocation. The
transmission is under power constraint, to be satisfied via the
precoder design and the amplification by a constant gain $A_0$, which
will be discussed later. The objective of DL optimization is to design
the preprocessing vectors or matrices, or so-called precoders, under
the constraints of APs' transmit power and/or other service
requirements.

When $s_m$ or $\tilde{\pmb{s}}_m$ for all $m\in\mathcal{M}$ are sent
over DL channels, it can be shown that the observation received by UE
$k$, $k\in\mathcal{K}$, can be expressed as
\begin{subequations}\label{eq:UCCF-32}
\begin{align}
y_k=&A_0\sum_{m\in\mathcal{M}}h_{mk}s_m+n_k\\
\tilde{\pmb{y}}_k=&A_0\sum_{m\in\mathcal{M}}\pmb{H}_{mk}^T\tilde{\pmb{s}}_m+\tilde{\pmb{n}}_k
\end{align}
\end{subequations}   
where $h_{mk}$ and $\pmb{H}_{mk}$ are the channels from UE $k$ to AP
$m$, and $n_k$ and $\tilde{\pmb{n}}_k$ are Gaussian noise distributed
with zero mean and a common variance $\sigma^2/2$ per dimension.

Substituting $s_m$ and $\tilde{\pmb{s}}_m$ from \eqref{eq:UCCF-31} respectively into \eqref{eq:UCCF-32}, it can be shown that  
\begin{subequations}\label{eq:UCCF-33}
\begin{align}
y_k=&A_0\pmb{h}_k^T\sum_{l\in\mathcal{K}}\pmb{p}_lx_l+n_k\\
\tilde{\pmb{y}}_k=&A_0\pmb{H}_k^T\sum_{l\in\mathcal{K}}\pmb{P}_l\pmb{\Phi}_l{\pmb{x}}_l+\tilde{\pmb{n}}_k
\end{align}
\end{subequations}   
where $\pmb{h}_k=[h_{1k},h_{2k},\ldots,h_{Mk}]^T$,
$\pmb{p}_l=[p_{1l},p_{2l},\ldots,p_{Ml}]^T$, and $\pmb{H}_k$ is
defined above \eqref{eq:UCCF-14}. $\pmb{p}_l\in\mathcal{C}^{M\times
  1}$ and $\pmb{P}_l\in\mathcal{C}^{MN\times N}$ are the vector and
matrix for preprocessing the signals sent to UE $k$. Corresponding to
$\pmb{H}_k$, we have
$\pmb{P}_k=\left[\pmb{P}_{1k}^T,\pmb{P}_{2k}^T,\ldots,\pmb{P}_{Mk}^T\right]^T$.

In \eqref{eq:UCCF-33}, the preprocessing vectors (matrices) can be
designed by exploiting the equivalency between the linear transmitter
preprocessing and linear receiver processing
(detection)~\cite{Lie-Liang-MC-CDMA-book,4686846,4212861,4212826,5278545}. In
other words, the preprocessing vector $\pmb{p}_k$ or matrix
$\pmb{P}_k$ can be obtained from the receiver processing vector
$\pmb{w}_k$ or matrix $\pmb{W}_k$ in the equivalent detection of
\begin{subequations}\label{eq:UCCF-34}
\begin{align}
 z_k=&\pmb{w}_k^H\left(\pmb{y}=\sum_{l\in\mathcal{K}}\pmb{h}_lx_l+\pmb{n}\right)\\
 \tilde{\pmb{z}}_k=&\pmb{W}_k^H\left(\tilde{\pmb{y}}=\sum_{l\in\mathcal{K}}\pmb{H}_l\pmb{\Phi}_l{\pmb{x}}_l+\tilde{\pmb{n}}\right)
\end{align}
\end{subequations}     
yielding $\pmb{p}_k=\sqrt{\Delta_k}\pmb{w}_k^*$ and
$\pmb{P}_k=\pmb{W}_k^*\pmb{\Delta}_k^{1/2}$, where $\{\Delta_k\}$ and
the diagonal matrices $\{\pmb{\Delta}_k\}$, joining with $A_0$,
account for the power-allocation to different UEs and also to the
symbols sent on different subcarriers of a UE. Hence, based on
\eqref{eq:UCCF-34}(a), we have
\begin{align}\label{eq:UCCF-35}
\pmb{p}_k=\sqrt{\Delta_k}\underbrace{\left(\sum_{l\in\mathcal{K}}\pmb{h}_l^*\pmb{h}_l^T+\sigma^2\pmb{I}_M\right)^{-1}}_{(\pmb{R}_y^*)^{-1}}\pmb{h}^*_k,~k\in\mathcal{K}
\end{align} 
in MMSE principle, where $\pmb{R}_y$ is the autocorrelation matrix of
$\pmb{y}$. The MMSE detection problem in \eqref{eq:UCCF-34}(b) has
been analyzed with the GMMSE detector in
Section~\ref{section-6G-4.5.1.1}. After ignoring the power-control
coefficients in \eqref{eq:UCCF-19}, we have
\begin{align}\label{eq:UCCF-36}
\pmb{P}_k=\underbrace{\left(\sum_{l\in\mathcal{K}}\pmb{H}_l^*\pmb{\Phi}_{l}\pmb{\Phi}_{l}^T\pmb{H}_l^T+\sigma^2\pmb{I}_{MN}\right)^{-1}}_{(\tilde{\pmb{R}}_y^*)^{-1}}\pmb{H}_k^*\pmb{\Delta}_k^{1/2},~~k\in\mathcal{K}
\end{align}
where $\tilde{\pmb{R}}_y$ is the autocorrelation matrix of $\tilde{\pmb{y}}$.

Note that, in \eqref{eq:UCCF-35} and \eqref{eq:UCCF-36}, the channel
vectors (or matrices) are depended on the large-scale propagation
pathloss and shadowing, as well as the small-scale fading.  Hence, the
transmitter preprocessing can naturally take their effect into
account. To achieve the best possible performance in terms of
mean-square error (MSE)\index{Mean-square error (MSE)}, a stronger
signal will be weighed by a bigger factor, while a stronger
interference draws more attention of the precoder for its suppression.

Furthermore, it is worth noting that, while the preprocessing vectors
$\{\pmb{p}_k\}$ or matrices $\{\pmb{P}_k\}$ are computed at CPU in a
global optimization way, they are however transmitted by the
distributed APs under their local constraints, such as, APs' transmit
power. Specifically for UE $k$, AP $m$ only transmits $p_{mk}$ in
$\pmb{p}_k$ or $\pmb{P}_{mk}$ in $\pmb{P}_k$. Therefore,
$\{\Delta_k\}$ in \eqref{eq:UCCF-35} or $\{\pmb{\Delta}_k\}$ in
\eqref{eq:UCCF-36} are just power-allocation
coefficients\index{Power-allocation!coefficient}. In order for
$\pmb{p}_k$ or $\pmb{P}_k$ to maintain its properties when it goes
through the channels to UE $k$, $\pmb{p}_k$ or $\pmb{P}_k$ must be
linearly amplified by the $M$ distributed APs. Specifically
considering $\pmb{p}_k$, this means that a same positive scalar $A_0$
should be multiplied on the $M$ different elements of $\pmb{p}_k$ by
the $M$ distributed APs, so as to protect $\pmb{p}_k$ from any
distortion.  Without any doubt, this is highly challenging in
practical implementation.

Upon substituting \eqref{eq:UCCF-35} into \eqref{eq:UCCF-33}(a), the SINR of UE $k$'s detection can be derived to be
\begin{align}\label{eq:UCCF-37}
\gamma_{k}\left(\{\Delta_k\}\right)=\frac{|\pmb{h}_k^T\pmb{p}_k|^2}{\sum_{l\neq k}|\pmb{h}_k^T\pmb{p}_l|^2+\sigma^2/A_0^2}
\end{align}
which is a function of the power-allocation coefficients
$\{\Delta_k\}$.  Similarly, when specifically considering the $i$th
symbol of UE $k$, from \eqref{eq:UCCF-33}(b), we can obtain the SINR
of
\begin{align}\label{eq:UCCF-38}
\gamma_{ki}\left(\{\pmb{\Delta}_k\},\{\pmb{\Phi}_l\}\right)=\frac{|\pmb{h}_{ki}^T\pmb{P}_k\pmb{\phi}_{ki}|^2}{\sum_{j\neq i}|\pmb{h}_{ki}^T\pmb{P}_k\pmb{\phi}_{kj}|^2+\sum_{l\neq k}\sum_{j=1}^{N_l}|\pmb{h}_{ki}^T\pmb{P}_l\pmb{\phi}_{lj}|^2+\sigma^2/A_0^2}
\end{align}
which is a function of both the subcarrier-allocation reflected by
$\{\pmb{\Phi}_l\}$ and the power-allocation interpreted by
$\{\pmb{\Delta}_k\}$.

Hence, the DL sum-rate\index{Sum-rate} of UCCF network is
\begin{subequations}\label{eq:UCCF-39}
\begin{align}
R\left(\{\Delta_k\}\right)=&\sum_{k\in\mathcal{K}}\log_2\left[1+\gamma_k\left(\{\Delta_k\}\right)\right]\\
R\left(\{\pmb{\Delta}_k\},\{\pmb{\Phi}_l\}\right)=&\sum_{k\in\mathcal{K}}\sum_{i=1}^{N_k}\log_2\left[1+\gamma_{ki}\left(\{\pmb{\Delta}_k\},\{\pmb{\Phi}_l\}\right)\right]
\end{align}
\end{subequations}

The above analysis assumes an idealized network, where each AP sends
signals to all UEs. In a practical UCCF network, usually only a few of
APs that a UE is associated with receive signals from and send signals
to the UE. Hence, \eqref{eq:UCCF-35} and \eqref{eq:UCCF-36} must be
modified. Accordingly, the operations may be summarized as follows.
\begin{enumerate} 

\item $\pmb{R}_y$ in \eqref{eq:UCCF-35} or $\tilde{\pmb{R}}_y$ in
  \eqref{eq:UCCF-36} can be estimated from the UL signals forwarded by
  APs to CPU. Alternatively, if APs only forward their estimated
  channels of the associated UEs to CPU, then, CPU computes an
  approximate for $\pmb{R}_y$ or for $\tilde{\pmb{R}}_y$ using the
  estimated channels by setting all the other unknown elements to
  zeros.

\item After $\pmb{R}_y$ or $\tilde{\pmb{R}}_y$ is obtained, CPU
  computes the preprocessing vector $\pmb{p}_k$ using
  \eqref{eq:UCCF-35} or matrix $\pmb{P}_k$ using \eqref{eq:UCCF-36}
  for all $k\in\mathcal{K}$, without considering the optimization of
  power-allocation coefficients. Note that in the calculation of
  $\pmb{p}_k$ or $\pmb{P}_k$, $\pmb{h}_k$ or $\pmb{H}_k$ only includes
  the estimated elements related to the APs that UE $k$ is associated
  with.

\item CPU optimizes the power-allocation coefficients of
  $\{\Delta_k\}$ or that in $\{\pmb{\Delta}_k\}$, which will be
  further discussed in Section~\ref{section-6G-4.6.2}.

\item CPU computes $s_m$ using \eqref{eq:UCCF-31}(a), or $\tilde{\pmb{s}}_m$ using \eqref{eq:UCCF-31}(b) for all $m\in\mathcal{M}$.

\item For $m=1,2,\ldots,M$, CPU sends $s_m$ or $\tilde{\pmb{s}}_m$ to
  AP $m$, where $s_m$ or $\tilde{\pmb{s}}_m$ is sent to DL after the
  same amplification by a gain $A_0$ informed by CPU.

\end{enumerate}

As above-mentioned, the TMMSE precoder designed by the central CPU but
used for signal transmission by distributed APs is challenging in
practical implementation. To ease the implementation challenges,
distributed preprocessing and transmission\index{Distributed
  preprocessing}\index{Distributed transmission} by APs may be
employed. Specifically considering the subcarrier-level precoding,
once the CSI is available for DL transmission, AP $m$ can send the
signals as
\begin{align}\label{eq:UCCF-41}
s_m=\sum_{k\in\mathcal{K}_m}\sqrt{P_{mk}}\left(\frac{h_{mk}^*}{|h_{mk}|}\right)x_k,~m\in\mathcal{M}
\end{align}
where $P_{mk}$ is AP $m$'s transmit power towards UE $k$. When all the
$M$ APs transmit signals synchronously and coherently, the received
signal by UE $k$ is
\begin{align}\label{eq:UCCF-42}
y_k=&\sum_{m\in\mathcal{M}_k}\sqrt{P_{mk}}|h_{mk}|x_k+\sum_{m\in\mathcal{M}_k}\sum_{l\in\mathcal{K}_m,l\neq k}\sqrt{P_{ml}}\left(\frac{h_{mk}h_{ml}^*}{|h_{ml}|}\right)x_l\nonumber\\
&+\sum_{m\in\bar{\mathcal{M}}_k}\sum_{l\in\mathcal{K}_m}\sqrt{P_{ml}}\left(\frac{h_{mk}h_{ml}^*}{|h_{ml}|}\right)x_l+n_k
\end{align}
where on the right-hand side, the first term is the desired signal,
the second term is the MUI generated by the UEs associated with the
APs that UE $k$ is also associated with, the third term is the MUI
from the APs other than that in $\mathcal{M}_k$, and $n_k$ is
noise. Explicitly, the signal received by UE $k$ experiences strong
interference, provided that UE $k$ shares some APs with other UEs.

Hence, additional resources are required to mitigate the interference
seen in \eqref{eq:UCCF-42}. One approach is to equip each AP with
multiple antennas, the number of which is bigger than the number of
UEs associated with it in the normal operational scenarios. In this
case, \eqref{eq:UCCF-41} can be modified to
\begin{align}\label{eq:UCCF-43}
\pmb{s}_m=\sum_{k\in\mathcal{K}_m}\sqrt{P_{mk}}\left(\frac{\pmb{p}_{mk}^*}{\sqrt{|\pmb{p}_{mk}|^2}}\right)x_k,~m\in\mathcal{M}
\end{align}
Assume $U(\geq |\mathcal{K}_m|~\forall m)$ antennas per AP. Then, in
\eqref{eq:UCCF-43},
$\pmb{p}'_{mk}={\pmb{p}_{mk}^*}/{\sqrt{|\pmb{p}_{mk}|^2}}$ is a
$U$-length preprocessing vector for transmitting $x_k$. Let the
channel vector from UE $k$ to AP $m$ be expressed as
$\pmb{h}_{mk}$. Let express
$\pmb{H}_m=[\pmb{h}_{m1},\pmb{h}_{m2},\ldots,\pmb{h}_{m|\mathcal{K}_m|}]$. Then,
in the principles of transmitter zero-forcing
(TZF)~\cite{Lie-Liang-MC-CDMA-book}\index{Transmitter zero-forcing
  (TZF)}, the preprocessing vector $\pmb{p}_{mk}$ for UE $k$ is the
$k$th column of
\begin{align}\label{eq:UCCF-44}
\pmb{P}_m=\pmb{H}_m^*\left(\pmb{H}_m^T\pmb{H}_m^*\right)^{-1}
\end{align}
Then, corresponding to \eqref{eq:UCCF-42}, the signal received by UE
$k$ is
\begin{align}\label{eq:UCCF-45}
y_k=&\sum_{m\in\mathcal{M}_k}\sqrt{P_{mk}}x_k+\sum_{m\in\bar{\mathcal{M}}_k}\sum_{l\in\mathcal{K}_m}\sqrt{P_{ml}}\left(\frac{\pmb{h}_{mk}^T\pmb{p}_{ml}^*}{\sqrt{|\pmb{p}_{ml}|^2}}\right)x_l+n_k
\end{align}
It shows that the signals sent by AP $m$ with $m\in\mathcal{M}_k$ do
not interfere UE $k$. UE $k$ still experiences MUI from the UEs that
do not share any APs with UE $k$. However, these interfering UEs as
well as the APs that these UEs are associated with should be located
relative far away from UE $k$. Consequently, their interference on UE
$k$ is insignificant.

Instead of TZF, TMMSE-based precoding can be implemented by the
distributed APs to transmit signals to their associated
UEs. Accordingly, the precoding vectors for AP $m$ to send signals to
its associated UEs can be obtained from the columns
of~\cite{Lie-Liang-MC-CDMA-book}
\begin{align}\label{eq:UCCF-46}
\pmb{P}_m=\pmb{H}_m^*\left(\pmb{H}_m^T\pmb{H}_m^*+\sigma^2\pmb{I}_{|\mathcal{K}_m|}\right)^{-1}
\end{align}
where $\sigma^2$ is noise variance. 

In the case that the noise variance $\sigma^2$ in \eqref{eq:UCCF-46}
is not available at APs, it can be replaced by a regulation parameter
$\rho$, which can be optimized to strike a trade-off between noise
mitigation and MUI
suppression~\cite{Lie-Liang-MC-CDMA-book}. Specifically, when
$\rho=0$, the precoder is reduced to the TZF precoder of
\eqref{eq:UCCF-44}, which may amplify noise in low SNR region, but is
capable of fully removing MUI. When $\rho=\sigma^2$, it achieves the
TMMSE precoding, which has the capability to suppress noise, while
possibly leaving a little MUI in the detection. Furthermore, when
$\rho\rightarrow\infty$, the precoder becomes a transmitter
matched-filtering (TMF)\index{Precoder!matched-filtering (MF)}
assisted precoder. This can be understood from \eqref{eq:UCCF-46} that
the precoding vectors are given by the columns of
$\pmb{P}_m=\pmb{H}_m^*$, when $\rho\rightarrow\infty$. As TMMSE, TMF
precoding has the capability to mitigate noise, but it treats MUI as
noise and hence, can hardly achieve a satisfactory MUI suppression
result, especially, in the high SNR region where MUI dominates.

\index{Downlink!precoding|)}
\subsection{Downlink Resource Optimization}\label{section-6G-4.6.2}

Corresponding to the problem of UL optimization, as described in
\eqref{eq:UCCF-RO31}, a general problem for the DL optimization to
maximize the sum-rate of an OFDM-UCCF network can be formulated
as\index{Optimization problem!maximize sum-rate}
\begin{subequations}\label{eq:UCCF-47}
\begin{align}
\{\zeta_{mk}^*\},&\{\delta_{nk}^*\},\{\Delta_{kn}^*\}\nonumber\\
&=\arg\max_{\{\zeta_{mk}\},\{\delta_{kn}\},\{\Delta_{kn}\}}\left\{\sum_{k\in\mathcal{K}}\sum_{n\in\mathcal{N}}\delta_{kn}\log_{2}\left(1+\gamma_{kn}\left(\{\zeta_{mk}\},\{\Delta_{kn}\}\right)\right)\right\}\\
s.t.~~& \zeta_{mk}\in\{0,1\},~\forall m\in\mathcal{M}, k\in\mathcal{K}\\
& \delta_{kn}\in\{0,1\},~\forall k\in\mathcal{K}, n\in\mathcal{N}\\
& \sum_{k\in\mathcal{K}}\sum_{n\in\mathcal{N}}\Delta_{kn}\leq 1\\
&\|\tilde{\pmb{s}}_m\|^2\leq P_m^{\max},~\forall m\in\mathcal{M}\\
&A_0:~|\tilde{\pmb{s}}_m(n)|^2\leq P_m^{\max}(n),~\forall m\in\mathcal{M}, n\in\mathcal{N}\\
&\sum_{n\in\mathcal{N}}\delta_{kn}\log_{2}\left(1+\gamma_{kn}\left(\{\zeta_{mk}\},\{\Delta_{kn}\}\right)\right)\geq R_k^{\min},~\forall k\in\mathcal{K}
\end{align}
\end{subequations}
where \eqref{eq:UCCF-47}(b), \eqref{eq:UCCF-47}(c) and
\eqref{eq:UCCF-47}(g) have the same meaning as
\eqref{eq:UCCF-RO31}(b), \eqref{eq:UCCF-RO31}(c) and
\eqref{eq:UCCF-RO31}(e), respectively, in \eqref{eq:UCCF-RO31},
\eqref{eq:UCCF-47}(e) is the total power constraint on an AP, while
\eqref{eq:UCCF-47}(f) guarantees that the precoding vectors computed
by CPU can be linearly amplified by $A_0$ and sent by the distributed
APs. Note that in \eqref{eq:UCCF-47}(f), $|\tilde{\pmb{s}}_m(n)|^2$ is
the power of the $n$th element of $\tilde{\pmb{s}}_m$, and
$P_m^{\max}(n)$ is the maximum power allowed to transmit this
element. This constraint means that a common amplification gain $A_0$
implemented by the distributed APs should satisfy the power
constraints of $|\tilde{\pmb{s}}_m(n)|^2\leq P_m^{\max}(n),~\forall
m\in\mathcal{M}, n\in\mathcal{N}$.  Finally, \eqref{eq:UCCF-47}(d)
imposes the constraint on the power-allocation
coefficients\index{Power-allocation!coefficient} assigned to the
precoding vectors.

Notice in \eqref{eq:UCCF-RO31} and \eqref{eq:UCCF-47} that the UE
association is considered independently in both UL and DL
optimization. In reality, it should be optimized by jointly
considering UL and DL. Neverthless, both \eqref{eq:UCCF-RO31} and
\eqref{eq:UCCF-47} are NP hard to solve, and the UE association is
usually optimized separately before carrying out the other UL and DL
optimization.

As the UL optimization, the DL optimization also needs to be divided
into some sub-problems of, such as, UE association,
subcarrier-allocation and power-allocation, which may be optimized in
succession by various optimization algorithms~\cite{9650567,9586055},
from classic methods~\cite{9931474} to the more advanced deep
learning and heterogeneous graph neural network~\cite{10153484}
methods. In comparison with UL optimization, in the UCCF networks, the
power-allocation in DL optimization is more challenging, as the
precoders are globally computed by CPU but locally implemented by the
distributed APs. To illustrate this further, let us consider the DL
optimization in the relatively simpler subcarrier-level DL
transmission scheme, which has the precoding vectors as shown in
\eqref{eq:UCCF-35}. The optimization problem corresponding to
\eqref{eq:UCCF-47} can be formulated as
\begin{subequations}\label{eq:UCCF-48}
\begin{align}
\{\zeta_{mk}^*\},\{\Delta_{k}^*\}=&\arg\max_{\{\zeta_{mk}\},\{\Delta_{k}\}}\left\{\sum_{k\in\mathcal{K}}\log_{2}\left(1+\gamma_{k}\left(\{\zeta_{mk}\},\{\Delta_{kn}\}\right)\right)\right\}\\
s.t.~~& \zeta_{mk}\in\{0,1\},~\forall m\in\mathcal{M}, k\in\mathcal{K}\\
& \sum_{k\in\mathcal{K}}\Delta_{k}\leq 1\\
&A_0:~|s_m|^2\leq P_m^{\max},~\forall m\in\mathcal{M}\\
&\log_{2}\left(1+\gamma_{k}\left(\{\zeta_{mk}\},\{\Delta_{k}\}\right)\right)\geq R_k^{\min},~\forall k\in\mathcal{K}
\end{align}
\end{subequations}
where \eqref{eq:UCCF-48}(c) imposes the constraints on the
optimization of power-allocation coefficients, while
\eqref{eq:UCCF-48}(d) explains that the common amplification gain
$A_0$ should allow the $M$ APs to simultaneously satisfy their power
constraints. When these conditions are met, $\{\pmb{p}_k\}$ in
\eqref{eq:UCCF-35} will be linearly amplified using a scalar gain
$A_0$ by the distributed APs. Hence, their properties and
relationships will be retained after the transmitted signals go
through the DL channels to UEs, yielding the performance as
expected. However, in practice, this is challenging to achieve. This
is because distributed APs are operated with separated amplifiers and
oscillators, which are hard to be controlled to operate with the same
gain and in a nearly ideal synchronization state.

\section{Concluding Remarks}\label{section-6G-4.7}

In this chapter, the principles of channel training, UL detection and
DL precoding in UCCF networks have been analyzed. Specifically, on
channel training, the UCCF networks operated with TDD, IBFD and MDD
have been discussed, showing that a large overhead may be required by
a TDD-based UCCF network for channel estimation. Furthermore, the
signal transmission in TDD-based UCCF networks experiences the channel
ageing\index{Channel ageing} problem, which may significantly degrade
performance. By contrast, owing to the simultaneous UL/DL
transmissions, IBFD- or MDD-aided UCCF systems are free from the
channel ageing problem, while the overhead for channel training can be
significantly reduced, when compared with the TDD-relied UCCF
systems. Between IBFD and MDD, the SI in IBFD-based UCCF systems may
limit the channel estimation to achieve high reliability, but MDD is
free from this limitation.

On UL detection, the GMMSE detection operated at CPU, LMMSE detection
operated at CPU or by the cooperative APs, and the APPA detection
carried out by the cooperative APs with the aid of CPU have been
analyzed. It can be expected that the APPA detector is a
high-efficiency detection scheme: it does not require strict
synchronization among APs, information exchange only occurs between
adjacent APs, there is little backhaul transmission between APs and
CPU, and it allows to achieve near-optimum performance.

On DL transmission, the principle of precoding has been analyzed, when
the transmitter preprocessing at both the subcarrier level and OFDM
symbol level is considered. Both the global precoder design by CPU and
the distributed precoder design by APs, as well as the implementation
issues of DL transmission and optimization have been addressed. The
analysis reveals that, while the precoder design for UCCF networks has
little difference from that for the classic MIMO, in particular, SDMA
or massive MIMO, systems, the implementation of DL transmission is
highly challenging. This challenge comes mainly from the requirement
that the centrally processed signals by CPU are transmitted by the
distributed APs, which may impose random distortions, resulting in
that the signals received from different APs by a UE are not
coherently added together. To ease this challenge, distributed AP
precoding may be an alternative, but extra resources, e.g., multiple
antennas, are required by APs to mitigate MUI.

Furthermore, in this chapter, both the UL and DL optimization issues
have been discussed, showing that the optimization, especially the DL
optimization, is more demand than that in the conventional BS centric
systems. For example, in the BS centric systems, BS carries out the DL
optimization under the constraint of its own transmit power. It
transmits the DL signals under the control of a same oscillator. By
contrast, in a UCCF network, the DL optimization may be done by CPU
but under the power constraints of distributed APs. Moreover, the
optimized signals by CPU are required to be linearly amplified by the
distributed APs and sent from them in nearly ideal synchronization,
which are difficult to implement in practice, when considering that
distributed APs are operated with independent amplifiers and
oscillators.

While the implementation of DL transmission is challenging in the UCCF
networks, where CPU carries out the centralized global optimization
but distributed APs are responsible for signal transmission, such
regularized transmission strategy does provide an interesting method
for CPU to send secrecy information to UEs. With this regard, first,
the information to UEs is secret to APs. This can be inferred, for
example, by \eqref{eq:UCCF-35} and \eqref{eq:UCCF-31}(a), where the
preprocessing vectors $\{\pmb{p}_k\}$ in \eqref{eq:UCCF-35} are
computed by CPU, but AP $m$ sends the combination of the $m$th
elements in $\{\pmb{p}_k\}$ and the data symbols in
$\mathcal{K}_m$. Hence, AP $m$ only knows the combination of the $m$th
elements in $\{\pmb{p}_k\}$ and the data symbols in $\mathcal{K}_m$,
not know the data and also the other elements in $\{\pmb{p}_k\}$.  As
shown in \eqref{eq:UCCF-35}, $\{\pmb{p}_k\}$ are dependent on the
channels between different UEs and different APs. Since AP $m$ only
knows the channels between it and its associated UEs, it should be
very difficult for AP $m$ to derive the information sent to its
associated UEs, needless to say the information sent to the other UEs.

Second, the information sent to a UE is secrete to the other UEs. This
can be conceived, for example, from \eqref{eq:UCCF-33}(a). When TZF or
TMMSE precoder is employed, UE $k$ is only capable of picking up its
own information.  The information sent to the other UEs is either
fully removed, when TZF precoding is employed, or mostly removed, when
TMMSE precoding is employed.

Third, the situation cannot be better for eavesdroppers. First,
eavesdroppers are hard to obtain the CSI between APs and UEs. Second,
the precoders are designed to achieve coherent receiving at the
locations of UEs. Since eavesdroppers usually have different locations
from UEs', it is impossible for them to attain the receiving effect as
\eqref{eq:UCCF-33}(a) for the legitimate UEs. Hence, the information
sent to UEs is normally secret to eavesdroppers. Perhaps, an active
eavesdropper may use a powerful receiver, such as, by employing
multiple receive antennas, to eavesdrop. In this case, an artificial
interference\index{Artificial interference} may be added to the
transmitted signal, in the form of
\begin{align}\label{eq:UCCF-49}
\pmb{s}=&\sqrt{\rho}\sum_{l\in\mathcal{K}}\pmb{p}_lx_l+\sqrt{1-\rho}\pmb{p}_In_I
\end{align}  
where $n_I$ is the random artificial interference, $\pmb{p}_I$ is
designed to be orthogonal with all the channel vectors from UEs to
APs, i.e., $\pmb{h}_k^T\pmb{p}_I=0,\forall k\in\mathcal{K}$, and
$\rho\in [0,1]$ controls the power assigned to transmit information
and artificial interference, respectively, which is an optimization
parameter.  Then, AP $m$ sends
\begin{align}\label{eq:UCCF-50}
s_m=&A_0\left(\sqrt{\rho}\sum_{l\in\mathcal{K}}p_{ml}x_l+\sqrt{1-\rho}p_{mI}n_I\right),~m\in\mathcal{M}
\end{align} 
The received signal by UE $k$ is
\begin{align}\label{eq:UCCF-51}
y_k=&A_0\sqrt{\rho}\pmb{h}_k^T\sum_{l\in\mathcal{K}}\pmb{p}_lx_l+n_k
\end{align}  
which, except the power scaling by $\rho$, is the same as
\eqref{eq:UCCF-33}(a). However, owing to the added artificial
interference $p_{mI}n_I$, the signals sent to UEs should become harder
to decode by the distributed APs and eavesdroppers.


\end{document}